\documentclass[sigconf,letterpaper]{acmart}

\renewcommand\footnotetextcopyrightpermission[1]{} 
\usepackage{amsmath,amssymb,amsfonts}
\usepackage{subfigure}
\usepackage{textcomp}
\usepackage{xcolor}
\usepackage{graphicx}
\usepackage{multirow}
\usepackage{textcomp}
\usepackage{tikz}
\usepackage{xcolor}
\usepackage{hyperref}
\usepackage{xspace}
\usepackage{listings}
\usepackage{bbding}
\usepackage{wasysym}
\usepackage{caption}
\usepackage{enumitem,kantlipsum}
\usepackage{textcomp}
\usepackage{xcolor}
\usepackage{soul}
\usepackage{interval}
\usepackage{pgfplots}
\usepackage{pifont}
\usepackage{amsmath}
\usepackage{amsfonts}
\usepackage{algorithm}
\usepackage{phonetic}
\usepackage{interval}
\usepackage{pgfplots}
\usepackage{tipa}
\usepackage{multicol}
\usepackage{multirow}
\usepackage{xspace}
\usepackage{xfrac}
\usepackage{pifont}
\usepackage{booktabs}

\newcommand{\cmark}{\ding{51}}%
\newcommand{\xmark}{\ding{55}}%
\usepackage[noend]{algpseudocode}

\newcommand{\supervoice}{\mbox{\textsc{SuperVoice}}\xspace}
\newcommand{\rev}[1]{{\color{black} #1}}

\definecolor{nickorange}{HTML}{03A60D}

\definecolor{nickorange}{HTML}{ffa500}

\definecolor{green}{HTML}{3049D4}

\definecolor{blue}{HTML}{0024ff}

\definecolor{myred}{RGB}{205 44 44}

\definecolor{myred}{RGB}{215 17 17}

\algrenewcommand\algorithmicforall{\textbf{foreach}}
\algrenewcommand\algorithmicindent{.8em}
\def\BibTeX{{\rm B\kern-.05em{\sc i\kern-.025em b}\kern-.08em
    T\kern-.1667em\lower.7ex\hbox{E}\kern-.125emX}}
\AtBeginDocument{%
  \providecommand\BibTeX{{%
    \normalfont B\kern-0.5em{\scshape i\kern-0.25em b}\kern-0.8em\TeX}}}

\begin{document}
\fancyhead{}
\title{\supervoice: Text-Independent Speaker Verification Using Ultrasound Energy in Human Speech}

\author{Hanqing Guo}
\email{guohanqi@msu.edu}
\affiliation{
  \institution{Michigan State University}
  \city{East Lansing}
  \state{MI}
  \country{USA}
  \postcode{48824}
}

\author{Qiben Yan}
\email{qyan@msu.edu}
\affiliation{
  \institution{Michigan State University}
  \city{East Lansing}
  \state{MI}
  \country{USA}
  \postcode{48824}
}

\author{Nikolay Ivanov}
\email{ivanovn1@msu.edu}
\affiliation{
  \institution{Michigan State University}
  \city{East Lansing}
  \state{MI}
  \country{USA}
  \postcode{48824}
}

\author{Ying Zhu}
\email{zhuying4@msu.edu}
\affiliation{
  \institution{Michigan State University}
  \city{East Lansing}
  \state{MI}
  \country{USA}
  \postcode{48824}
}

\author{Li Xiao}
\email{lxiao@cse.msu.edu}
\affiliation{
  \institution{Michigan State University}
  \city{East Lansing}
  \state{MI}
  \country{USA}
  \postcode{48824}
}

\author{Eric J. Hunter}
\email{ejhunter@msu.edu}
\affiliation{
  \institution{Michigan State University}
  \city{East Lansing}
  \state{MI}
  \country{USA}
  \postcode{48824}
}


\begin{abstract}
Voice-activated systems are integrated into a variety of desktop, mobile, and Internet-of-Things (IoT) devices. However, voice spoofing attacks, such as impersonation and replay attacks, in which malicious attackers synthesize the voice of a victim or simply replay it, have brought growing security concerns. Existing speaker verification techniques distinguish individual speakers via the spectrographic features extracted from an audible frequency range of voice commands. However, they often have high error rates and/or long delays. In this paper, we explore a new direction of human voice research by scrutinizing the unique characteristics of human speech at the ultrasound frequency band.
Our research indicates that the high-frequency ultrasound components (e.g. speech fricatives) from 20 to 48 kHz can significantly enhance the security and accuracy of speaker verification.
We propose a speaker verification system, \supervoice that uses a two-stream DNN architecture with a feature fusion mechanism to generate distinctive speaker models.
To test the system, we create a speech dataset with 12 hours of audio (8,950 voice samples) from 127 participants. In addition, we create a second spoofed voice dataset to evaluate its  security. In order to balance between controlled recordings and real-world applications, the audio recordings are collected from two quiet rooms by 8 different recording devices, including 7 smartphones and an ultrasound microphone.
Our evaluation shows that \supervoice achieves 0.58\% equal error rate in the speaker verification task, which reduces the best equal error rate of the existing systems by 86.1\%. \supervoice only takes 120 ms for testing an incoming utterance, outperforming all existing speaker verification systems. Moreover, within 91 ms processing time, \supervoice achieves 0\% equal error rate in detecting replay attacks launched by 5 different loudspeakers. Finally, we demonstrate that \supervoice can be used in retail smartphones by integrating an off-the-shelf ultrasound microphone. 
\end{abstract}

\ccsdesc[500]{Security and privacy~Biometrics}

\keywords{Voice Authentication; Ultrasound; Speaker Verification}

\maketitle
\section{Introduction}
\label{sec:introduction}
Modern devices increasingly adopt biometric technologies for user authentication. 
Among various types of human biometrics, such as fingerprint, facial, iris, etc., voice biometric demonstrates great benefits in its high usability, convenience, and security. 
\emph{Speaker Verification (SV)} systems commonly use voice biometrics to automatically accept or reject a voice input based on the speaker models
stored on the smart devices or cloud. Nowadays, all the popular voice assistants, such as Siri, Alexa, and Google Assistant, have integrated SV algorithms for certain wake words (e.g., ``Hey, Siri'', ``Ok, Google''). 

A more appealing approach, called \emph{text-independent speaker verification}, could accurately and efficiently verify arbitrary utterances from a target speaker based on a limited set of enrolled sentences. Recently, security researchers have demonstrated 
the susceptibility of SV systems to voice mimicry attacks and replay attacks, where the attackers imitate victims' voices or record/replay them to bypass the SV systems~\cite{janicki2016assessment,kinnunen2017asvspoof,wu2015spoofing,alegre2012vulnerability}. As the number of sensitive  applications (e.g., banking~\cite{usbank}) of voice assistants is growing, practical SV systems aim to achieve not only high accuracy in text-independent speaker verification but also high efficacy in defending spoofing attacks under a limited time budget.

\begin{figure}[t]
    \centering
    \includegraphics[width=3.35in]{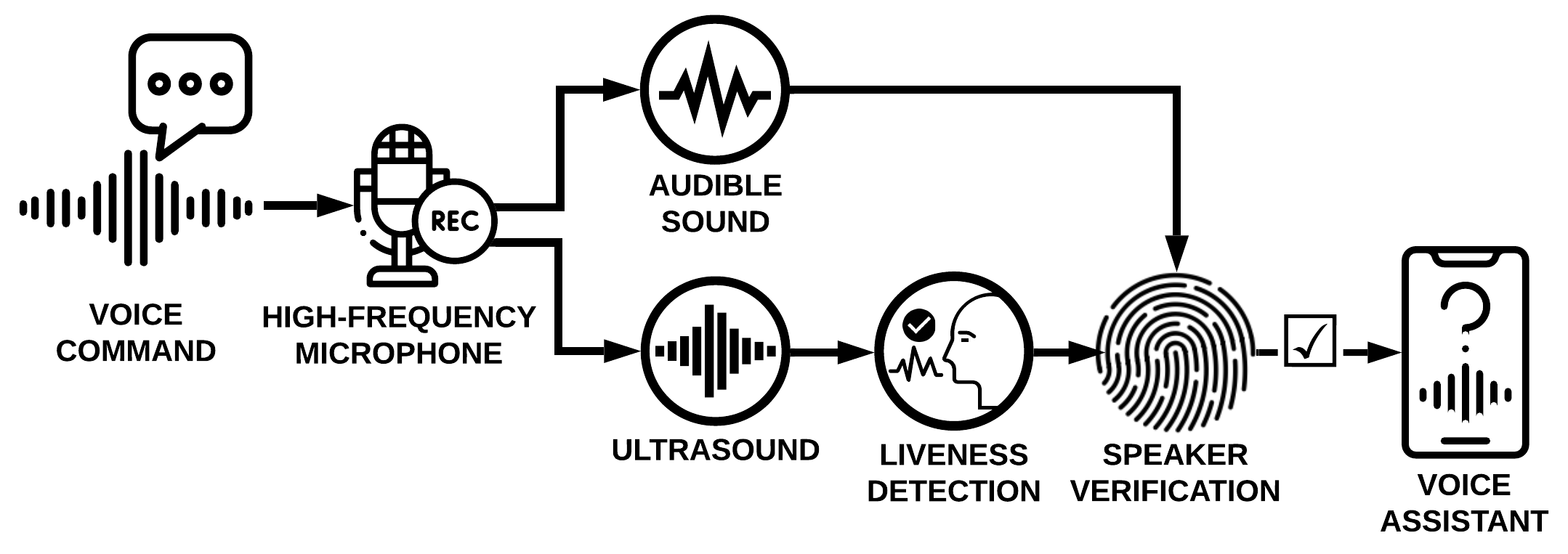}
    \vspace{-13pt}
    \caption{The workflow of \supervoice.}
    \label{fig:supervoice-overview}
    \vspace{-10pt}
\end{figure}

Recent SV studies have explored the distinctive vocal or non-vocal features such as phoneme position~\cite{zhang2016voicelive}, cumulative spectrum~\cite{ahmedvoid}, mouth motion~\cite{meng2018wivo, zhang2017hearing}, body vibration~\cite{feng2017continuous}, and sound field~\cite{yan2019catcher}. Based on these features, conventional machine learning models have been used to generate speaker models, including correlation (CORR), support vector machine (SVM), Gaussian mixture models (GMM), etc.
Meanwhile, deep neural network (DNN) based SV systems use robust neural networks for building speaker models with the prototypical features (e.g., waveform~\cite{ravanelli2018speaker, hoshen2015speech, sainath2015learning}, spectrogram~\cite{heigold2016end, nagrani2020voxceleb, wan2018generalized}, and MFCC (Mel-Frequency Cepstral Coefficients)~\cite{team2017hey}). 
As summarized in Table~\ref{tab:comparison1}, most of the existing SV systems cannot simultaneously achieve effective speaker verification and defense against spoofing attacks~\cite{zhang2016voicelive,zhang2017hearing,ahmedvoid, meng2018wivo}, while others have limitations in their usability, e.g., with the requirement of wearing extra devices~\cite{feng2017continuous}, staying at the same positions as the enrollment phase~\cite{yan2019catcher}, etc. Moreover, their discovered vocal or non-vocal features cannot be \emph{transferred} across different speaker models. Although existing DNN-based SV systems~\cite{heigold2016end, nagrani2020voxceleb, wan2018generalized, team2017hey, ravanelli2018speaker} do not deal with rigid features, they tend to yield relatively high error rates due to the lack of speaker representative features. 

\noindent\textbf{Motivation:}
In this research, we aim to explore the ultrasound energy in human speech to enhance the accuracy and security of text-independent speaker verification. 
More specifically, we investigate the unique properties of the \emph{human speech in the human-inaudible ultrasound frequency band} (i.e., frequencies greater than $20$ kHz). 
High-frequency ultrasound components in human speech present several unique properties: first, they are imperceptible by humans but can be captured by an ultrasound microphone;
second, individual human speakers can produce ultrasound waves with distinct characteristics, determined by the unique shape of the speech production system and the particular use (e.g. timing) of the system.
Recent attacks towards voice assistants, such as DolphinAttack~\cite{zhang2017dolphinattack} and SurfingAttack~\cite{yan2020surfingattack}, leveraged the inaudible ultrasound signals to inject commands into voice assistants. Here, we take a \emph{reversed} approach: rather than utilizing the ultrasound signals for attack, we take advantage of the unique properties of the high-frequency audio spectrum for  defense,
with the goal of offering a more robust and accurate SV system.

We propose \supervoice, a robust and secure text-independent SV system, which is applicable to commodity mobile devices equipped with an ultrasound microphone.
\supervoice analyzes an incoming voice command to the device microphone, as shown in Fig.~\ref{fig:supervoice-overview}. The audio spectrum of the voice command includes both the audible (below $20$ kHz) and ultrasound (above 20 kHz) frequency components. \supervoice then processes these components to extract both the low-frequency and high-frequency feature representations using a liveness detection module and a \emph{two-stream DNN architecture}. These features are fused to a second level of classifier to generate or match a speaker embedding for the speaker verification purpose.

\noindent\textbf{Challenges:} The design of \supervoice faces 3 critical challenges.
\emph{i) How to ascertain that the ultrasound feature can represent speaker's voiceprint?} Prior acoustic studies show the evidences that high-frequency energy (from 8-16 kHz) contains useful features to identify an individual speaker~\cite{schwartz1968identification, jongman2000acoustic, hayakawa1994text}.
However, none of them focuses on the ultrasound frequency band above 20 kHz. The existing feature engineering techniques such as LPCC (Linear Prediction Cepstral Coefficients), Filter banks, and MFCC cannot be directly applied in high-frequency data, as they are designed for the narrowband speech data (below 8 kHz). To better utilize the ultrasound features, we design signal processing techniques to extract the unique characteristics from the ultrasound components. 
\emph{ii) How to use the ultrasound features to detect replay attacks that involve multiple playback devices?} Since the attackers can use different devices (e.g., smartphones, ultrasonic microphone, and ultrasonic speaker) to record and replay the voice signals, it is challenging to design an liveness detection method to cope with different attack devices with varied signal characteristics. 
\emph{iii) How to design a neural network structure to integrate the ultrasound features?} Since the speech production theory of the low-frequency features and high-frequency features are very different, the integration of both features is particularly challenging. We design a two-stream DNN structure with convolutional filters to process and integrate the ultrasound features. 

\noindent\textbf{Contributions:}
To the best of our knowledge, we are \emph{the first} to 
prove that \emph{ultrasound components (20 $\sim$ 48 kHz)} in human speech can be used to enhance the accuracy, robustness, and security of the SV systems. We demonstrate that the ultrasound components are model-agnostic by integrating them into multiple SV models, all of which achieve enhanced performance in the SV tasks. Surprisingly, the ultrasound components in human speech have been largely neglected prior to this work~\cite{0888Jayant1984}. 
\begin{table}
\scriptsize
    \caption{\supervoice in comparison with other SV systems.}
    \label{tab:comparison1}
    \centering
        \begin{tabular}{p{1.3cm}p{1.2cm}p{0.8cm}p{0.9cm}p{0.5cm}p{0.5cm}}
          \toprule[1.5pt]
          \small{\textbf{System}} & \small{\textbf{Feature}} &
          \small{\textbf{Model}} &
          \small{\textbf{Text Indep.}} & \small{\textbf{Secu-rity}} & \small{\textbf{Trans-fer}}\\
          \bottomrule[1.5pt]
          VoiceLive~\cite{zhang2016voicelive} & Phoneme & CORR & \xmark & \cmark & \xmark \\
          \hline
          VoiceGes.~\cite{zhang2017hearing} & Mouth & CORR & \xmark & \cmark & \xmark \\
          \hline
          WiVo~\cite{meng2018wivo} & Mouth & CORR &\xmark & \cmark & \xmark \\
          \hline
          VAuth~\cite{feng2017continuous} & Body & CORR &\cmark & \cmark & \xmark \\
          \hline
           Void~\cite{ahmedvoid} & Cum. Spec & SVM & \cmark & \cmark & \xmark \\
          \hline
          CaField~\cite{yan2019catcher} & Sound field & GMM &\cmark & \cmark & \xmark \\
          \hline
          TE2E~\cite{heigold2016end} & Spectrum & CNN &\cmark & \xmark & \cmark \\
          \hline
          GE2E~\cite{wan2018generalized} & Spectrum & CNN &\cmark & \xmark & \cmark \\
          \hline
          Siri~\cite{team2017hey} & MFCC & RNN & \cmark & \xmark & \cmark \\
          \hline
          SincNet~\cite{ravanelli2018speaker} & Waveform & CNN& \cmark & \xmark & \cmark \\
          \hline
          VGGVox~\cite{nagrani2020voxceleb} & Spectrum & CNN &\cmark & \xmark & \cmark\\
          \hline
          \textbf{\supervoice} & \textbf{Ultrasound} & \textbf{CNN} & {\cmark} & {\cmark} & {\cmark} \\
      \bottomrule[1.5pt]
      \multicolumn{5}{l}{
    \vspace{-20pt}} \\
        \end{tabular}
\end{table}
In summary, our paper makes the following contributions:
\begin{itemize}
    \item We demonstrate that human speech does include ultrasound components, and those components can help distinguish among different human speakers. Moreover, the ultrasound components in speech signals can be utilized to identify spoofing attacks by measuring the signals' cumulative energy levels at the high-frequency range.
    \item We design \supervoice, a speaker verification and spoofing detection system. By incorporating a two-stream neural network structure with time-frequency spectrogram filters and feature fusion mechanism, \supervoice achieves high accuracy in text-independent speaker verification. 

    \item We launch a human voice study and collect two datasets for speaker verification and spoofing detection. We recruit 127 participants and record 8,950 audios by 8 different smart devices, including 7 smartphones and an ultrasound microphone. We also replay 500 audio samples to construct a spoofed voice dataset with 5 playback devices. In total, our datasets involve 127 participants with a total of 12 hours audios.
    We make our datasets publicly available at 
    \textbf{\url{https://supervoiceapp.github.io/}}.

    \item We evaluate the performance of \supervoice and compare it against other SV systems. The result shows that \supervoice achieves 0.58\% equal error rate (EER) in speaker verification, which improves the EER of the top-performing SV system by 86.1\%. Remarkably, it only takes 120 ms to test an incoming utterance. Moreover, \supervoice achieves 0\% EER with 91 ms processing time for liveness detection, outperforming all existing approaches. The two-week longevity experiment demonstrates that \supervoice is suitable for a long-term use, and its performance is barely impacted by the changes in distances and angles.
\end{itemize}
\section{Background}
\label{sec:background}
\subsection{Threat Model}
We consider \emph{voice spoofing attack}, which is a malicious attempt to impersonate a genuine speaker in order to execute an unauthorized command on a voice assistant. Three most popular types of voice spoofing attacks include \textit{replay}, \textit{synthesis}, and \textit{conversion}~\cite{wu2015spoofing}. In a replay attack, the adversary records the legitimate command uttered by a genuine speaker, and replays this command later. The synthesis attack uses the text-to-speech (TTS) generation to create artificial voice commands acceptable by a voice assistant. The conversion attack  converts an existing voice command into a different one that can bypass speaker verification. 
To provide effective countermeasures against voice spoofing attacks, this research aims to develop an end-to-end SV system that can perform both liveness detection and speaker verification.

\subsection{Can Humans Produce Ultrasound?}\label{sec:make_ultra}
\begin{figure}
    \centering
    \includegraphics[width=1.8in]{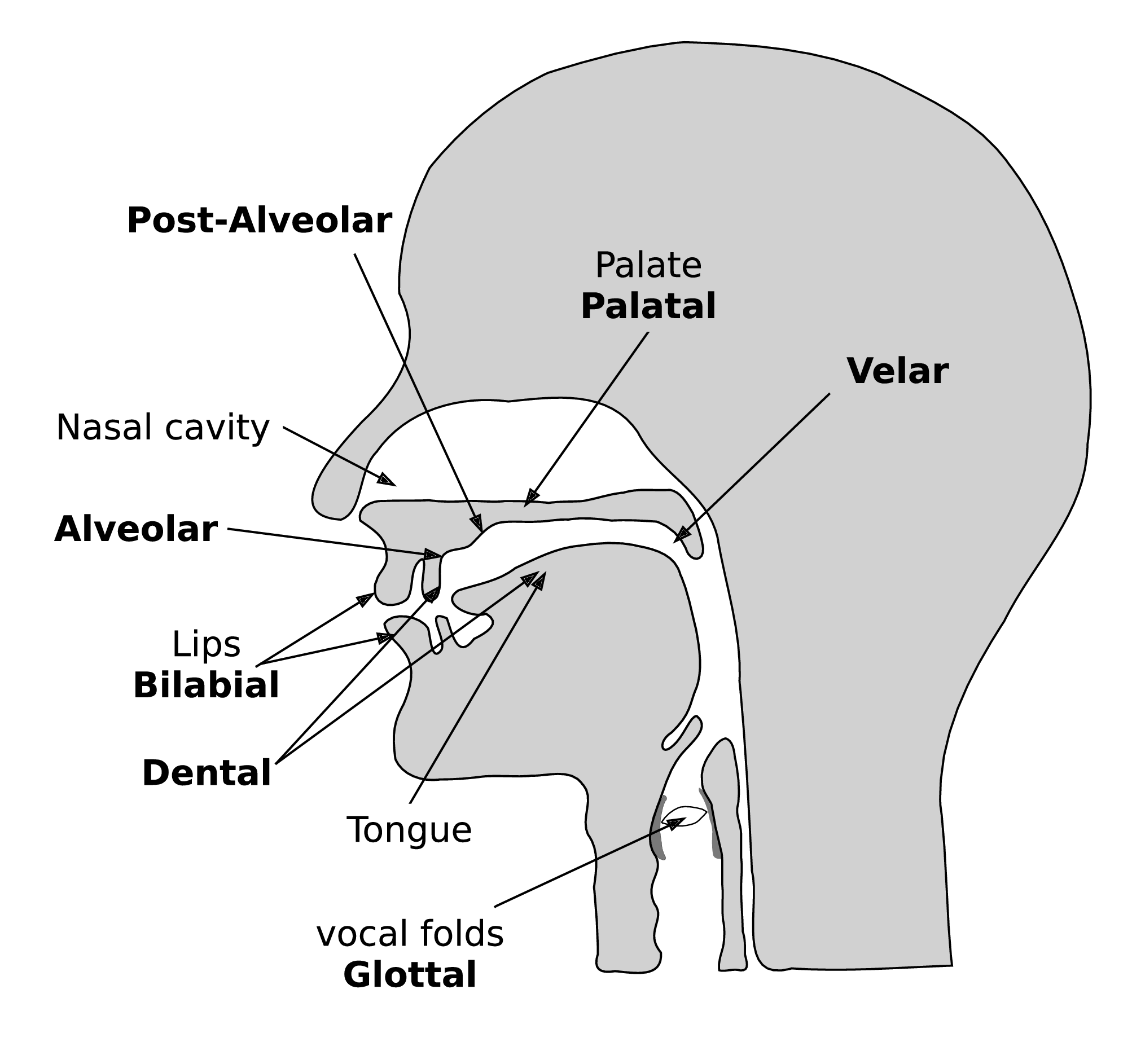}
    \vspace{-10pt}
    \caption{Human's vocal tract and place of articulation.}
    \label{fig:vocal-tract}
    \vspace{-15pt}
\end{figure}

The sounds in human speech are commonly divided into vowels and consonants: the vowels are produced when the air steadily flows through the vocal tract above the larynx (see Fig.~\ref{fig:vocal-tract}), while the consonants are more transient in nature and are produced when the flow of air is partially restricted or completely stopped at the vocal fold. 
The consonants are characterized by \emph{voicing}, \emph{place of articulation}, and \emph{manner of articulation}~\cite{jakobson1951preliminaries,chomsky1968sound}. Voicing refers to the inclusion of vocal fold vibration as sound source which are quasi-steady and generally harmonic in nature, and the place of articulation represents the location of constriction in the vocal tract which usually results in a highly transient noise.  
The manner of articulation describes how a sound is altered by the manipulation of airstream flows from the lungs. 
More specifically, 

when two speech organs narrow the airstream to cause friction to occur as it passes through, \emph{Fricatives} are produced.
If the airstream is stopped and then released, \emph{Stop} or \emph{Affricate} is produced. 

\begin{figure*}[t]
\centering     
\subfigure[Spectrum of a given phrase ]{\label{fig:whole_spec}\includegraphics[width=42mm]{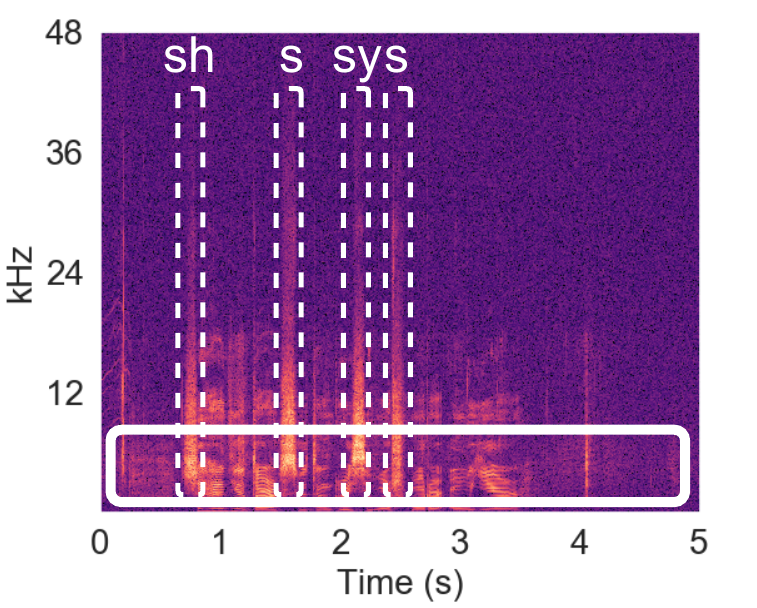}}
\subfigure[Spectrum of LFE components]{\label{fig:detail_spec}\includegraphics[width=42mm]{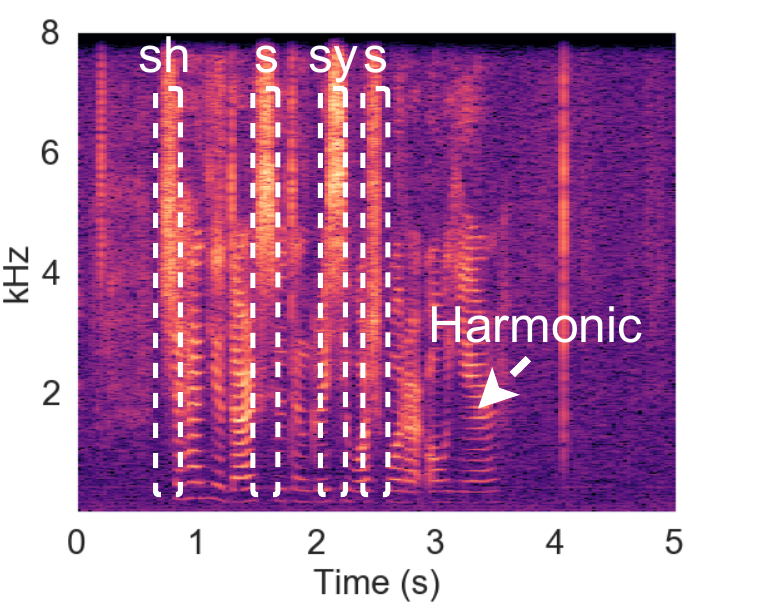}}
\subfigure[Voice spectrum comparison ]{\label{fig:phoneme_comp}\includegraphics[width=42mm]{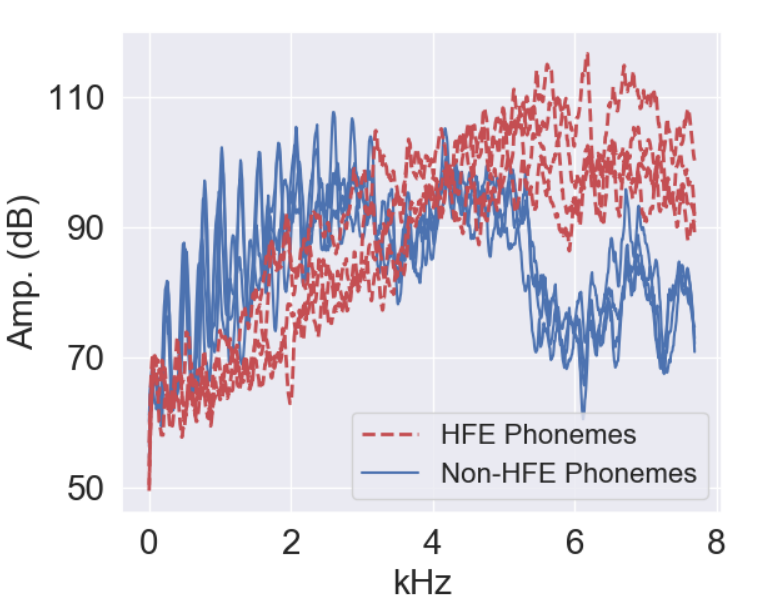}}
\subfigure[Replayed audio spectrum]{\label{fig:replay_demo}\includegraphics[width=42mm]{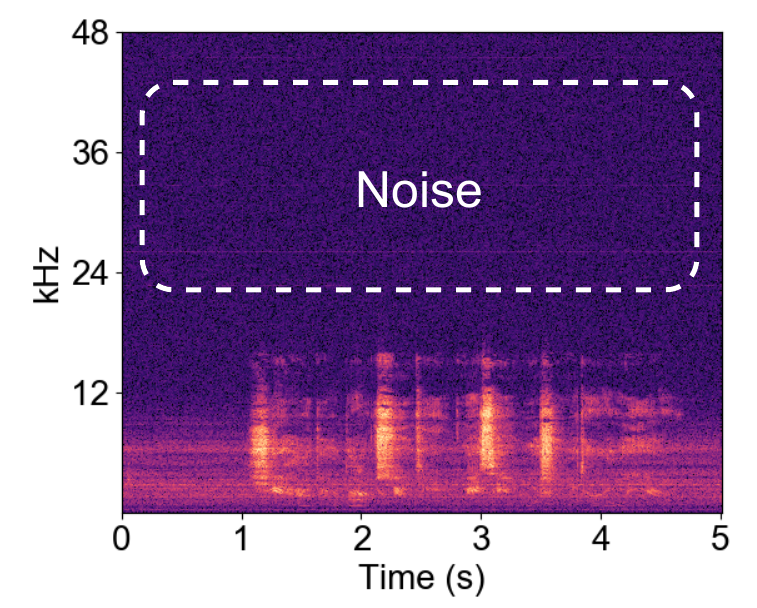}}
\vspace{-10pt}
\caption{Observation of high frequency energy (HFE) and low frequency energy (LFE) of the phrase \textit{``\underline{S}he had your dark \underline{s}ui\textbf{t} in grea\underline{sy} wa\underline{s}h water all year''} uttered by a human speaker.}
\vspace{-7pt}
\label{fig:observations}
\end{figure*}

Particularly, the Stop, Affricate, and Fricative consonants
are known to exhibit 
\emph{high frequency energy (HFE)}, since the airstream is deformed by articulations. 
In this work, \emph{we aim to scrutinize this under-explored and largely neglected phenomenon in human speech, i.e.,
the consonants carry high energy in the human-inaudible ultrasound frequency range}. We perform experiments to validate that a human speech production generates energy in the ultrasound spectrum during a normal utterance, primarily within speech components such as the Stop, Affricate, and Fricatives.
Fig.~\ref{fig:whole_spec} shows the human voice frequency spectra sensed by an ultrasound microphone, in which a significant portion of the acoustic energy is observed beyond $20$ kHz. In this study, we are the \emph{first} to show that the acoustic energy beyond $20$ kHz (i.e., ultrasound voice components) plays an important role in the 
speaker verification task and offers an effective and economical solution for liveness detection. 

\subsection{Can Ultrasound Components Improve SV Performance?}\label{sec:necessity}

Carefully examining Fig.~\ref{fig:whole_spec}, we find that HFE is produced by certain phonemes (marked by dashed rectangles), such as /sh/, /s/, /sy/, and /s/ within the particular phrase. Fig.~\ref{fig:detail_spec} shows the low frequency spectrum of these phonemes, from which we can see that the phonemes with HFE exhibit less energy below 2 kHz compared with other phonemes. 

\noindent\textbf{Remark 1:} The phonemes with HFE may lack low-frequency energy (LFE). This phenomenon implies that the traditional LFE-based SV systems may not be able to capture sufficient characteristics of the phonemes with HFE. 

Everyone has a unique vocal track, which leads to a distinctive pitch, formant, and harmonic pattern on the traditional sonic voice spectrum. 
The modern SV models follow this principle to identify voiceprint by modeling the energy distribution pattern from the LFE spectrum. 
Fig.~\ref{fig:phoneme_comp} shows an obvious difference in the voice spectrum between the phonemes with HFE and the ones without HFE. Therefore, by capturing the unique high-frequency characteristics in the phonemes with HFE, the ultrasound components may help boost the performance of the text-independent SV systems. 

The most common audio sampling rate of a recorder or loudspeaker is 44.1 (or 48) kHz. Due to the Nyquist theorem, any acoustic energy beyond 22.1 (or 24) kHz will be  discarded as shown in Fig.~\ref{fig:replay_demo}. Even though some recorders and loudspeakers have higher sampling rate, their frequency responses tend not to be as flat across a wide frequency band as the human speech. 

\noindent\textbf{Remark 2:} Typical replay attack using loudspeakers could not produce the ultrasound energy. Therefore, the ultrasound energy in human speech can be used to quickly identify loudspeaker-based spoofing attacks.

In this following sections, we present our new discoveries on the specific features of human voice components, which become the core elements of \supervoice. 
By conducting a preliminary study on the high-frequency ultrasound components in human speech, 
we lay the foundation for the rest of this work. The preliminary study aims to answer the following four complementary research questions:

\begin{itemize}
    \item \textbf{RQ1:} \emph{Can the ultrasound components in human speech reflect the speaker identity?}
    
    \item \textbf{RQ2:} \emph{How consistent are the ultrasound features for each speaker over different speech contents?}
    
    \item \textbf{RQ3:} \emph{How distinctive are the ultrasound features for each individual speaker?}
    
    \item \textbf{RQ4:} \emph{Can the ultrasound components help determine the liveness of the audio input?}
    
\end{itemize}

\begin{figure*}[t]
\centering     
\subfigure[Energy of diff. sentences from the same speaker ]{\label{fig:diff_sents_same_spk}\includegraphics[width=42mm]{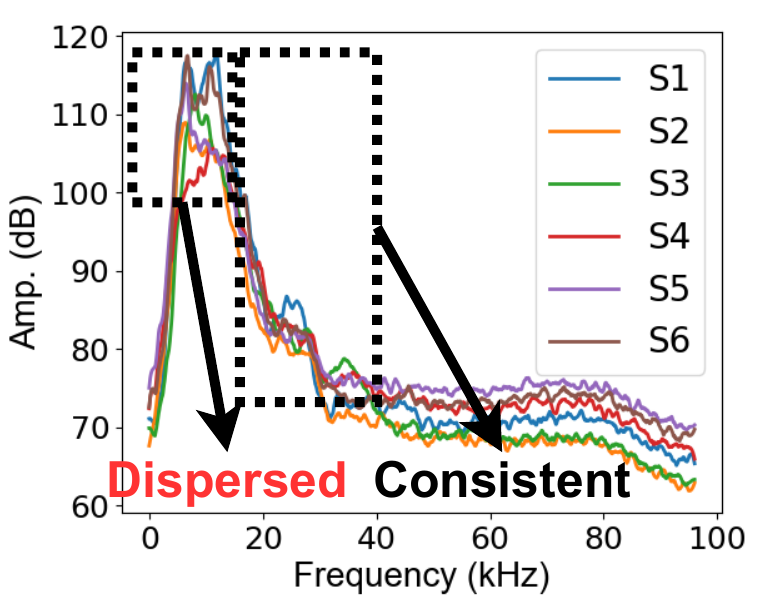}}
\subfigure[Variance of energy w.r.t. frequencies]{\label{fig:variance_same_spk}\includegraphics[width=42mm]{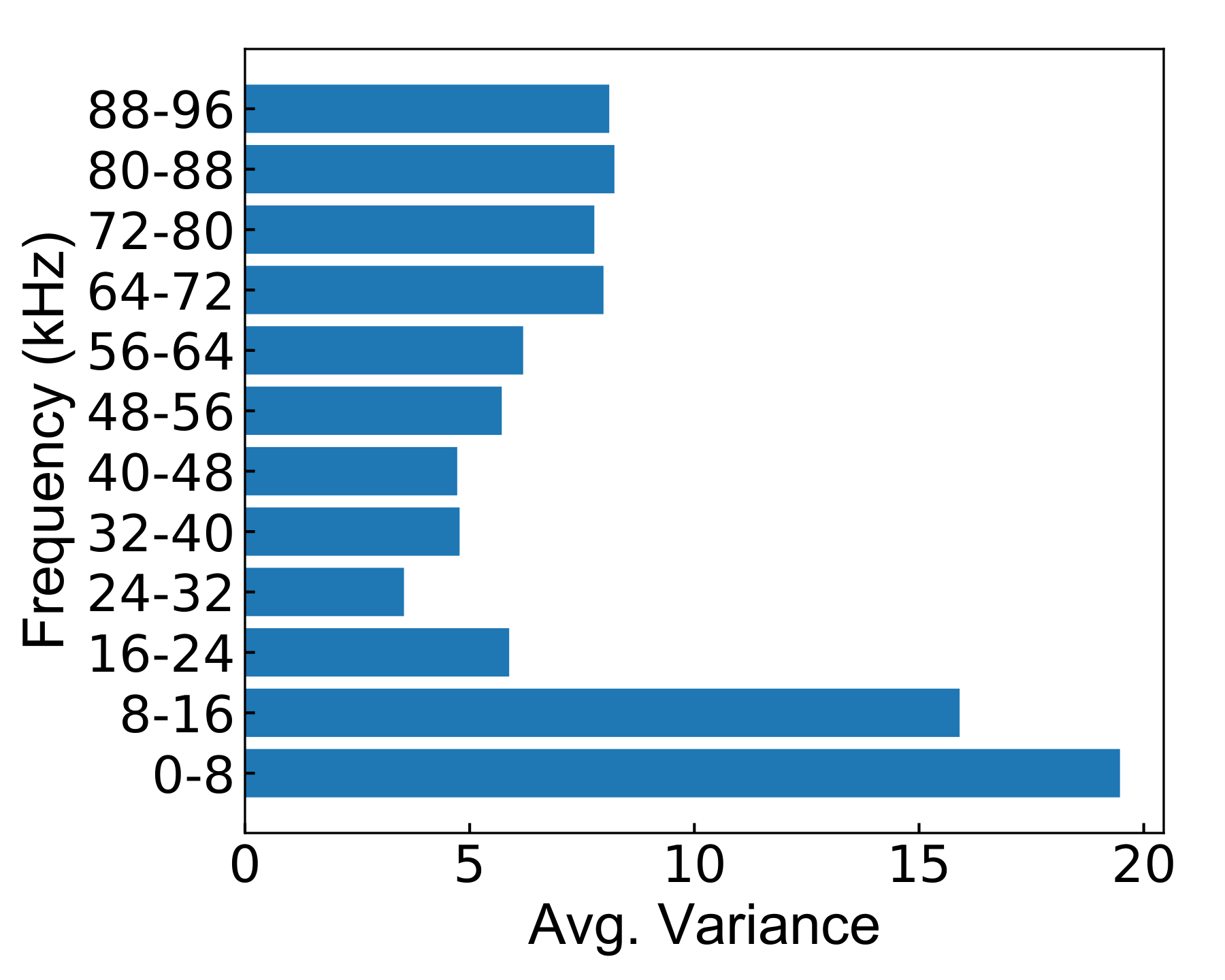}}
\subfigure[Energy of the same sentences from diff. speakers ]{\label{fig:same_sents_diff_spk}\includegraphics[width=42mm]{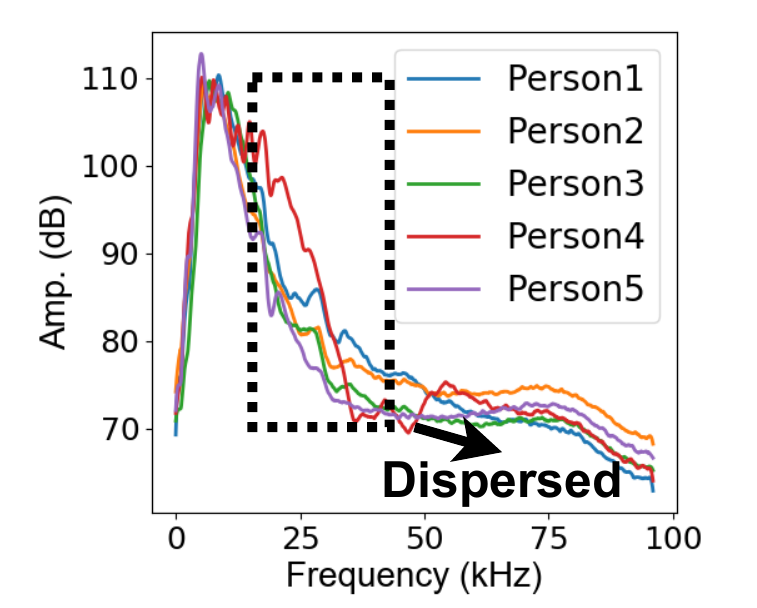}}
\subfigure[Variance of energy w.r.t. frequencies]{\label{fig:variance_diff_spk}\includegraphics[width=42mm]{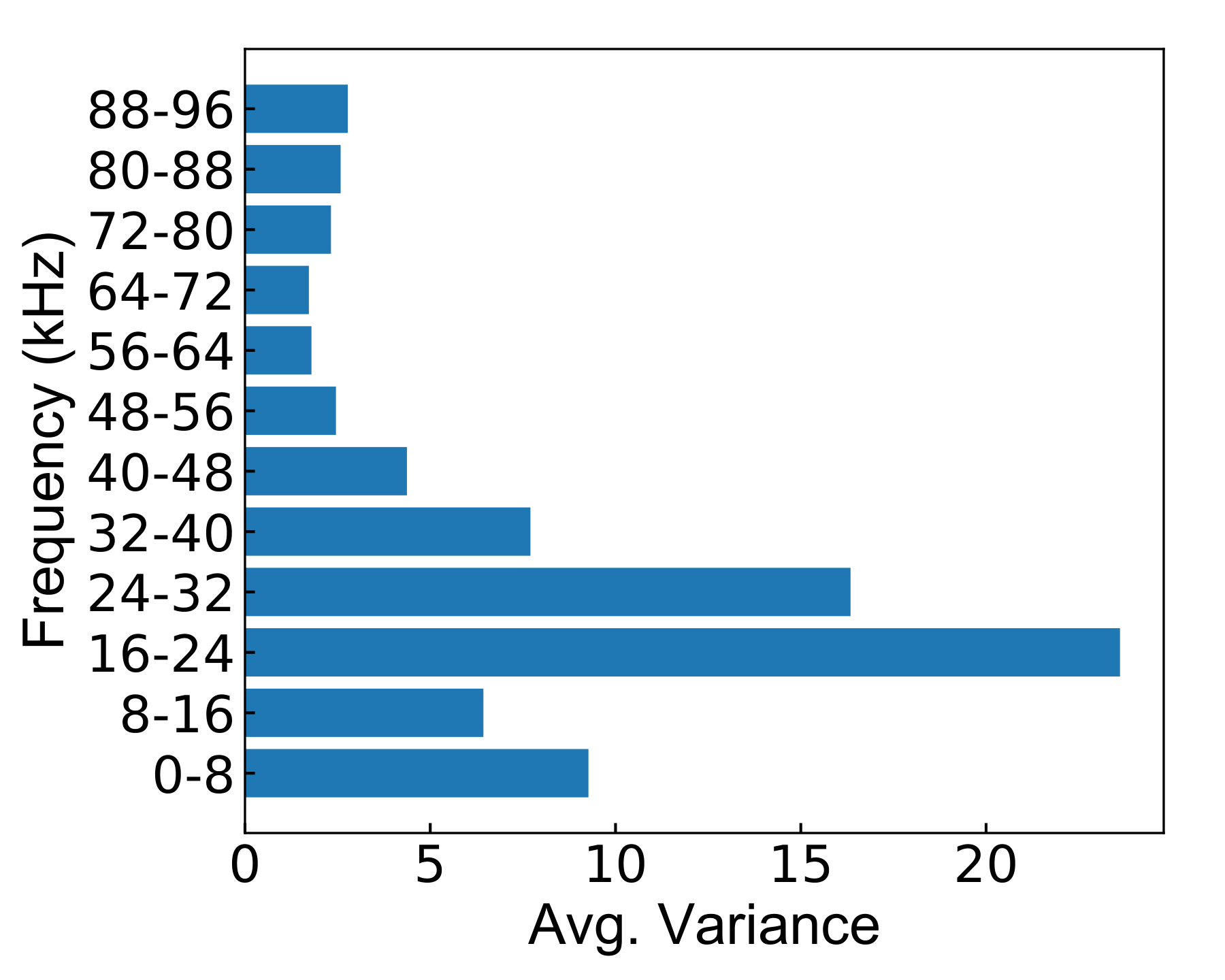}}
\vspace{-10pt}
\caption{Ultrasound energies of different sentences spoken by different speakers.}
\vspace{-7pt}
\end{figure*}

\subsection{Ultrasound Components and Speaker Identity}
To answer \textbf{RQ1}, we conduct a theoretical analysis based on the principle of human speech. Generally, the production of speech can be divided into two separate elements: 1) sources of sound such as the larynx, and 2) filters that modify the sources such as the vocal tract. Different from the vowels that only use voicing source, the consonants coordinate three sources: \emph{frication, aspiration}, and \emph{voicing}. Moreover, vowels are produced by relatively open vocal tract, while consonants are produced by constrictions in the vocal tract, as explained in Section~\ref{sec:make_ultra}. 
Specifically, the production of consonants involves more sources, faster changes in articulators, changes in the shape of the vocal tract, and more complicated articulation changes such as the movement of the tongue, lips, and jaw.  
As a result, the consonants naturally produce a more diverse set of frequency components, which extends to the ultrasound frequency range. 
Clearly, the uniqueness of the consonant pronunciation depends on a human's vocal tract shape, nasal cavity, oral cavity structure, and the lip and tongue shapes. Among all consonants, we focus on the \emph{Stop, Affricate, and Fricative} consonants, since they produce high-frequency components with a significantly higher energy level (see Fig.~\ref{fig:whole_spec}). 

\subsection{Consistency of Ultrasound Components}
To address \textbf{RQ2}, we design an experiment to evaluate whether the ultrasound frequency components are consistent across different speech contents. 
Conceptually, the \emph{ultrasound component} refers to the speech component with a non-trivial energy above 20 kHz. We first identify the high-energy ultrasound components in an utterance by computing the Short-time Fourier transform (STFT) spectrum of the voice input. The STFT uses a Hann window of length 10 ms, hop length of 2 ms, and FFT size of 2,048 points under 192 kHz sampling rate, which results in 93.75 Hz ($\frac{192,000}{2,048}$) frequency resolution. Suppose an utterance is divided into $N$ frames as each lasts $T$ in time. We consider the top $M$ frames with the highest cumulative energy above 20 kHz as the frames that contain ultrasound components. Based on empirical observations, $M$ is configured as $100$ in this paper. 

However, existing studies have demonstrated that STFT spectrum of different phonemes present notable deviations across certain frequency ranges~\cite{tabain2001variability, monson2014perceptual}. This indicates that the impact of speech contents could pose a challenge for text-independent SV scenarios. To address this challenge, we calculate the \emph{long term average (LTA)} of the energies of ultrasound components, and the LTA is more stable within the time frame $T$, expressed as follows:
\vspace{-5pt}
\begin{equation}
S_{LTA}(f)=\frac{1}{M}\sum_{t=1}^{M}S(f, t),
 \label{eq:lta}
 \vspace{-5pt}
\end{equation}
where $M$ is the number of frames that contain high-frequency ultrasound components, $S(f, t)$ is the STFT spectrogram at frequency $f$ and frame $t$, and $t$ is the frame index  within $T$. 
The spectrum averaging techniques such as LTA have been used to compare the properties of acoustic signals from random speech data~\cite{yan2019catcher}. In essence, LTA can help reduce the impact of different phonemes on the speaker profile. Here, we ask one volunteer to read the sentences S1-S6 (refer to the website \url{https://supervoiceapp.github.io}), 
and collect the spectrogram data to compute $S_{LTA}$. The results in Fig.~\ref{fig:diff_sents_same_spk} show that $S_{LTA}$ remains consistent within the frequency range between 16-48 kHz across different speech contents. The variance of $S_{LTA}$ is shown in Fig.~\ref{fig:variance_same_spk}. It is worth noting that $S_{LTA}$ (adapted for low frequency) varies significantly within the low frequency range between 0-16 kHz, which further corroborates that LTA of ultrasound components can be used to improve the performance of SV systems. 
\subsection{Distinctiveness of Ultrasound}
Next, we aim to address \textbf{RQ3}, i.e., whether the ultrasound features from human speech
are unique to each speaker, given that each speaker's vocal tract is unique. 
Prior to answering this research question, we formalize the ultrasound voiceprint for each speaker. The creation of a voiceprint typically involves training with multiple sentences to achieve a reliable and robust voiceprint. Suppose the enrollment dataset is $\mathbb{D}$. The ultrasound voiceprint $P$ is defined as:
\vspace{-5pt}
\begin{equation}
    P = \frac{1}{|\mathbb{D}|}\sum_{s\in \mathbb{D}} S_{LTA}(s),
    \vspace{-5pt}
    \label{equ:person}
\end{equation}
where $s$ denotes the sentence index, and $S_{LTA}(s)$ is the LTA of the ultrasound energy within the sentence $s$. 
The ultrasound voiceprint represents the average energy distributions of multiple enrollment sentences.

To evaluate the capability of $P$ in distinguishing among speakers, we enroll the voice of five volunteers (3 males and 2 females) 
and analyze the distinctiveness of $P$. 
The results in Fig.~\ref{fig:same_sents_diff_spk} demonstrate noticeable variations in the ultrasound energy in the range of 16-48 kHz. 
Fig.~\ref{fig:variance_diff_spk} further indicates that the ultrasound components from different speakers vary the most at the frequency range of 16-32 kHz. 

\subsection{Ultrasound for Liveness Detection}

The aforementioned experiments show that human voice possesses ultrasound components, but the digital loudspeaker generally cannot produce highly distinctive ultrasounds with a high energy. The sound spectrogram produced by a digital loudspeaker is limited by its  Analog Digital Converter (ADC) sampling rate, low-pass filters, amplifier, and other loudspeaker hardware. We demonstrate this phenomenon in Fig.~\ref{fig:whole_spec} and Fig.~\ref{fig:replay_demo}, where the former shows genuine human utterance has ultrasound components, while the latter spectrogram of a loudspeaker does not contain HFE in the high-frequency band. Therefore, \textbf{RQ4} can be addressed by measuring the ultrasound energy in audios. For high-end recorders and loudspeakers that support higher sampling rates, we demonstrate the effectiveness of our design in Section~\ref{sec:evaluation}.

\noindent\textbf{In summary,} we demonstrate the ultrasound components contain speaker identify information. We show that the ultrasound voiceprints based on ultrasound components are consistent across different speech contents. They are  distinctive for each speaker, and they can be used for liveness detection to enhance the security of SV systems.

\section{\supervoice Design}
\label{sec:design}
In this section, by leveraging the discriminative ability of the ultrasound components in the human voice, we introduce \supervoice to perform liveness detection and speaker verification simultaneously. 
We first extract the ultrasound features from the voice signals. 
Then, the ultrasound features are embedded into a two-stream DNN architecture to produce speaker embeddings via the integration of both the low-frequency and high-frequency voice components.

\begin{figure}[t]
    \centering
    \includegraphics[width=3.2in]{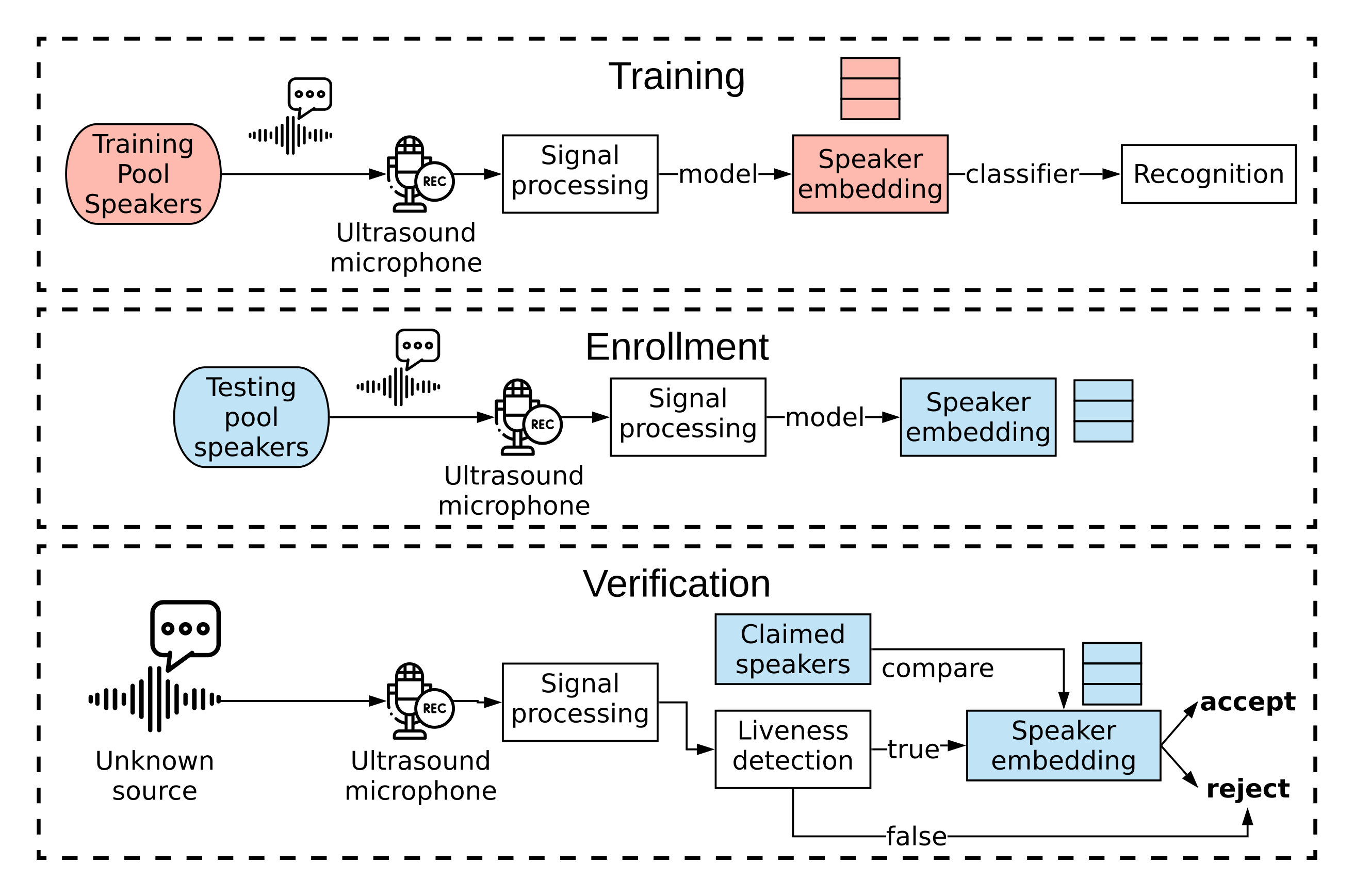}
    \vspace{-10pt}
    \caption{\supervoice's operational workflow.}
    \label{fig:modular}
    \vspace{-15pt}
\end{figure}

\subsection{Overview}
The goal of \supervoice is to utilize the high-frequency components of the human voice to enhance the speaker verification performance while identifying and defending against spoofing attacks. To achieve this goal, \supervoice includes 3 main stages: 1) \emph{Model Training}, 2) \emph{Speaker Enrollment}, and 3) \emph{Speaker Verification}. Fig.~\ref{fig:modular} shows the operational workflow of \supervoice. 

During the Model Training stage, \supervoice first learns how to extract effective features (speaker embeddings) from the voice data of speakers in the training pool. In Speaker Enrollment, the target speakers are required to register their voices in the system, based on which \supervoice will generate and store the speaker embedding as the speaker's unique voiceprint. Finally, in the Speaker Verification stage, \supervoice first conducts the liveness detection to ensure the audio source is spoken by a human speaker, and then verifies the speaker identity by measuring the similarity with the claimed speaker embeddings.
At every stage, a \emph{Signal Processing} module (see Appendix~\ref{sec:sp}) is applied to convert the raw signals to fit the input shape of the model. The processed data are then fed into the \emph{liveness detection} module and the \emph{two-stream DNN network} for speaker verification. 
\begin{figure*}
\centering
  \includegraphics[width=6in]{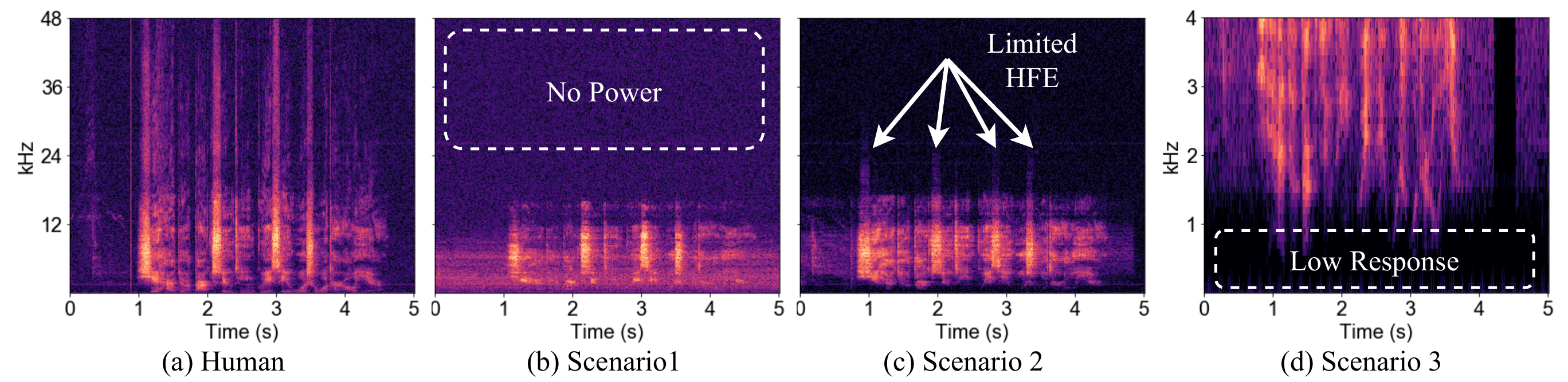}
  \vspace{-10pt}
  \caption{Spectrograms from different replay attackers.}
  \vspace{-7pt}
  \label{fig:live1}

\end{figure*}

\subsection{Liveness Detection}
\label{sec:live}
The liveness detection module is designed to differentiate between human voice and spoofed audio by utilizing cumulative spectral energy of captured audio frames with ultrasonic components. We consider three attack scenarios based on different attackers' capabilities.

\noindent\textbf{Scenario 1. Attackers record and replay with commercial devices:}
Since most of the traditional recording devices 1) do not support microphones that produce high-frequency response; 2) do not have an ADC capable of producing audios with a high sampling rate; and 3) often apply a low-pass filter behind the microphone logic to remove high-frequency components --- they are unable to acquire a full spectrum of human voice. An ultrasound microphone, on the other hand, can capture a wider spectrum of human voice (see Fig.~\ref{fig:live1}a), including the ultrasound components up to 48 kHz. The digital components of a loudspeaker usually have a sampling rate at or below 48 kHz. Therefore, the replayed audio will not carry any high-frequency ultrasound components as opposed to the genuine human voice (Fig.~\ref{fig:live1}b). As a result, the captured ultrasound components in human voice provide a unique opportunity for developing accurate and efficient liveness detection without heavy computation.

\noindent\textbf{Scenario 2. Attackers record with high-end microphones and replay with commercial speakers:} Let us consider an attacker, who uses a high-end microphones (e.g., a microphone with high sampling rate, high-resolution ADC, and wide frequency response) to eavesdrop a victim's voice, and replays it with a commercial speaker such as  smartphones or high-quality speakers. In such a scenario, the replayed audio will still carry limited HFE due to the cutoff frequency of commercial speakers, as shown in Fig.~\ref{fig:live1}c. In comparison with Fig.~\ref{fig:live1}a, the lack of HFE in Fig.~\ref{fig:live1}c constitutes a unique feature of the replay attacks. 

\noindent\textbf{Scenario 3. Attackers record with high-end microphones and replay with ultrasound speakers:} The attackers can also be equipped with high-end microphones and professional ultrasound speakers. In this scenario, although the spectrogram of the replayed audio carries HFE, it possesses limited LFE, as shown in Fig.~\ref{fig:live1}d. 

The energy difference between the replayed audio and genuine human voice is evident: the former has nearly zero energy below 1 kHz, while the later presents an intact spectrum.

Based on our observations in Fig.~\ref{fig:live1}, we leverage the cumulative spectral energy of the frames with ultrasonic components and design an \emph{accurate}, \emph{efficient}, and \emph{lightweight} liveness detector to identify if the audio source comes from a loudspeaker or a genuine human speaker. The detector relies on the normalized cumulative energy $S_{p}$ in different frequency ranges, as defined below:
\begin{equation}
    S_{p}(f) = \sum_{t\in M} S(f, t) -  \overline{\sum_{f}\sum_{t\in T} S(f, t)},
    \label{eq:spower}
\end{equation}
where $S$ is the STFT spectrogram, $t$ is the index of frames, $T$ is the total number of frames, and $M$ is the number of frames with ultrasonic components.  The first term of the right-hand side summarizes the energies for all the frames with ultrasonic components, and the second term is used for normalization and noise reduction. 

To defend against the attacks in Scenarios 1 and 2, we define $R_1$ as the ratio of the ultrasonic energy over the entire spectrum as follows:
\begin{equation}
    R_1 = \frac{\sum_{f=low_{1}} ^ {high_{1}} S_{p}(f)}{\sum_{f=0} ^ {high_{1}} S_{p}(f)}.
    \label{eq:spoof1}
\end{equation}
The numerator is composed of the normalized accumulative energy on the high-frequency band (from $low_1$ Hz to $high_1$ Hz), while the denominator uses the energy of the entire spectrum (up to $high_1$ Hz). In this paper, $low_1$ and  $high_1$ are set as 24 and 48 kHz, respectively. 
Typically, a legitimate human voice will yield a positive value of $R_1$, since its HFE is greater than the average energy of the entire spectrum (see Fig.~\ref{fig:live1}a). 
In contrast, a replay attacker with a commercial speaker will yield a negative $R_1$. 

For Scenario 3 in which the attacker possesses a professional ultrasound microphone and high-end loudspeaker, we propose $R_2$ to examine the proportion of LFE over all frequency bands as follows:
\begin{equation}
    R_2 = \frac{\sum_{f=0} ^ {low_{2}} S_{p}(f)}{\sum_{f=0} ^ {high_{2}} S_{p}(f)}. 
    \label{eq:spoof2}
\end{equation}
The normalized accumulative energy below 1 kHz is supposed to be negative for replayed audio, since it has lower energy as shown in the dotted frame in Fig.~\ref{fig:live1}d with a dark color. For instance, we set $low_2$ as 1 kHz and $high_2$ as 4 kHz. 

By integrating $R_1$ and $R_2$, we consider a voice input as belonging to a genuine human if it satisfies the $(R_1 > 0) \land (R_2 > 0)$ condition.
Otherwise, it will be classified as a replayed audio.

\begin{figure*}
\centering
  \includegraphics[width=6.5in]{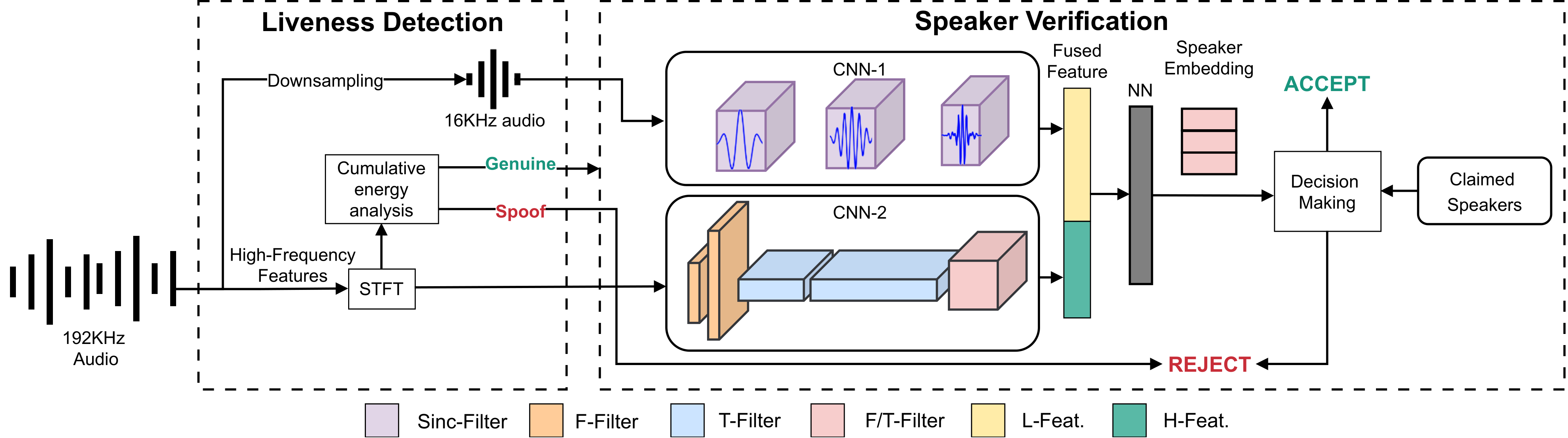}
  \caption{\supervoice system architecture (L-Feat. represents low-frequency feature embedding; H-Feat. represents high-frequency feature embedding).}
  \vspace{-7pt}
  \label{fig:model}
\end{figure*}

\subsection{Two-Stream DNN Model}
After performing the liveness detection, \supervoice begins processing the genuine human speech to verify the speaker identity. 
For speaker verification, we design a two-stream DNN model to better integrate ultrasound features to improve the SV performance. 

Almost all the prior SV studies consider low-frequency audios
below 8 kHz, because the typical voice characteristics such as pitch and formants only exist in low frequency range below 8 kHz. However,
we observe that the spectrum features above $8$ kHz can indeed contribute to the speaker verification. Thus, the question we aim to address in this section is: \emph{how to embed the high-frequency features into the speaker model?}

\noindent \textbf{DNN System Design:}
Typical machine learning based SV systems use the Neural Network (NN) to obtain feature embeddings~\cite{heigold2016end,variani2014deep,chen2015locally}. 
Followed by the feature embedding network, a classifier will perform the classification based on the extracted feature embeddings.
\supervoice follows such a structure, i.e., the first level of networks conducts feature embedding, while the second level performs the classification. 

\noindent \textbf{Feature Fusion:}
Different from the typical machine learning based SV system, \supervoice contains two streams of DNN models: one performs feature embedding using the low-frequency components, and the other one embeds high-frequency features. These features will be fused together to construct one coherent feature vector, and then fed into a classifier to produce a unique speaker embedding for every enrolled speaker. 

\vspace{-5pt}
\subsubsection{System Architecture}
The overall \supervoice architecture is presented in Fig.~\ref{fig:model}, which is comprised of three NNs:
CNN-1, CNN-2, and an NN classifier.

\noindent \textbf{CNN-1:}  CNN has been widely used in image recognition tasks, which applies convolutional computation to shrink the input size to obtain a high-level representation of high-dimensional features~\cite{zeiler2014visualizing}. We feed the downsampled raw audio containing low-frequency components into the CNN to obtain a low-frequency feature vector. Inspired by SincNet~\cite{ravanelli2018speaker}, we use Sinc-filters to simulate the behavior of band-pass filters, and add two one-dimensional convolutional layers to further compress the low-frequency feature space.

\noindent \textbf{CNN-2:} In the second data stream, CNN-2 is designed to embed high-frequency voice components. Since the existing CNNs, such as VGGNet~\cite{simonyan2014very}, ResNet~\cite{he2016deep}, Inception~\cite{szegedy2015going}, are designed for image classification, they apply 
multiple convolutional layers with different shapes (mostly squared shape) of filters to capture the unique characteristics of different object shapes in the image. As opposed to the image which consists of pixels, the spectrogram data possesses both time and frequency components. Here, we design a new CNN architecture with three complementary time-frequency convolutional filters to extract the HFE distribution and phoneme-level high-frequency representations.

\noindent \textbf{F-Filters:} The purpose of these frequency-domain filters (F-Filters) is to extract the HFE distribution $S(f)$ at the frequency domain.
We design a sequence of vertical shaped filters to convolve the high-frequency spectrogram. 
The size of F-Filter is determined by the range of frequencies involved in the convolutional computation.
Based on the observation that HFE distribution can be used as the speaker voiceprint (see Fig.~\ref{fig:diff_sents_same_spk}, \ref{fig:same_sents_diff_spk}), in order to extract a finer-grained energy distribution with a higher frequency resolution across the frequency range, we construct 
64 F-Filters, whose size is $9\times$1 with dilation 1$\times$1 and 2$\times$1. 
As a result, the filters cover the frequency range from $9\cdot 93.75=843.75$ Hz to $9\cdot 2\cdot 93.75=1,687.5$ Hz.

\noindent \textbf{T-Filter:} Two time-domain filters (T-Filter) are designed to learn the high-frequency phoneme-level representation. 
The T-Filter covers a time duration, which is shorter than a phoneme length to ensure that the convolution process can occur within a single phoneme. 
The time-domain resolution can be computed by $hop_{STFT}/192~kHz\approx 2.7$ ms. After applying the 64 1$\times$9 T-Filters that is dilated by 1$\times$1 and 1$\times$2, the convolution computation covers the time-domain resolution between $9\cdot 2.7=24.3$ ms and $9\cdot 2.7\cdot2=48.6$ ms. Since $48$ ms is shorter than typical phonemes, the time-domain frames can represent the detailed information from a single phoneme.

\noindent \textbf{F/T-Filter:}  At the final stage of CNN-2, we design a sequence of square filters (F/T-Filter) with the size of $5\times5$ to convolve both time-domain and frequency-domain features concurrently. F/T-Filter merges the extracted high-frequency characteristics from both the time and frequency domains, in order to yield a more representative ultrasound energy distribution for a particular speaker. 

\noindent \textbf{NN classifier:} Finally, the NN classifier takes the fused features that are concatenated by the output feature vectors of the CNN-1 and CNN-2 and compresses them into a desired speaker embedding dimension. Here, we use a fully connected layer as the NN classifier. 
More detailed neural network parameter settings can be found in Appendix~\ref{appendix:neural}.

\noindent\textbf{Speaker Embedding Generation and Matching:}
The speaker embedding is generated by the NN, as shown in Fig.~\ref{fig:model}. The NN is essentially a fully connected layer, which maps the fused feature vector to a speaker embedding vector. Please refer to Appendix~\ref{appen:feature} for the details of feature alignment. 

When the speaker model produces the speaker embedding based on the given audio source, \supervoice  will compare the cosine distance with the existing speaker embeddings, which have been generated during the enrollment stage. Every sentence spoken by an authorized speaker will produce a representative speaker embedding. 
For example, if speaker A enrolls three sentences into \supervoice, the model will generate three embeddings for A. 
To accept the audio as belonging to speaker A, the average cosine similarity between the tested speaker embedding and the enrolled one should be greater than a similarity threshold $\gamma$ as shown below:
\vspace{-10pt}
\begin{equation}
decision =
\begin{cases}
  accept, & similarity \ge \gamma \\
  reject, & similarity < \gamma
      \label{eq:decision}
\end{cases}
\end{equation}
where $similarity = \sum_{i = 0}^{N} cos(emb_i,emb)/N$, $N$ is number of enrolled audios for the speaker, $emb_i$ is the $i$\textsuperscript{th} speaker embedding, and $emb$ is the speaker embedding to be verified.

\vspace{-5pt}
\subsubsection{Model Training/Testing}
It is noteworthy that although the purpose of the NN models is to extract the speaker embedding, they operate differently in three stages (see Fig.~\ref{fig:modular}).
The model will learn how to extract the most representative speaker embeddings via training with a speaker recognition task. It means that the output of NN will connect to another fully-connected layer that maps the result dimension from speaker embedding to the number of classes in the speaker recognition task. For example, the model will predict a speaker label for the test utterance, and then refine the network parameters via the backpropagation of losses. In Speaker Enrollment stage, however, the model simply loads the set of parameters that achieve the best results in speaker recognition task, and then extracts the speaker embeddings for the enrolled speakers.

\section{Evaluation}
\label{sec:evaluation}
In this section, we evaluate the performance of \supervoice on spoofing detection and speaker verification, i.e., how well \supervoice can verify a claimed speaker and reject a spoofed audio or a stranger's voice.
Furthermore, we 
integrate the high-frequency features extracted by \supervoice into existing SV models to show the transferability of high-frequency features in enhancing different types of SV models. To have a fair evaluation, we collect several speech datasets as listed in Section~\ref{sec:data_collect}. 
Our experiments are conducted on a desktop with Intel i7-7700k CPUs, 64GB RAM, and NVIDIA 1080Ti GPU, running 64-bit Ubuntu 18.04 LTS operating system. The model complexity and time consumption are measured in such a hardware configuration.
\begin{figure}[!t]
\centering
\includegraphics[height=1.8in]{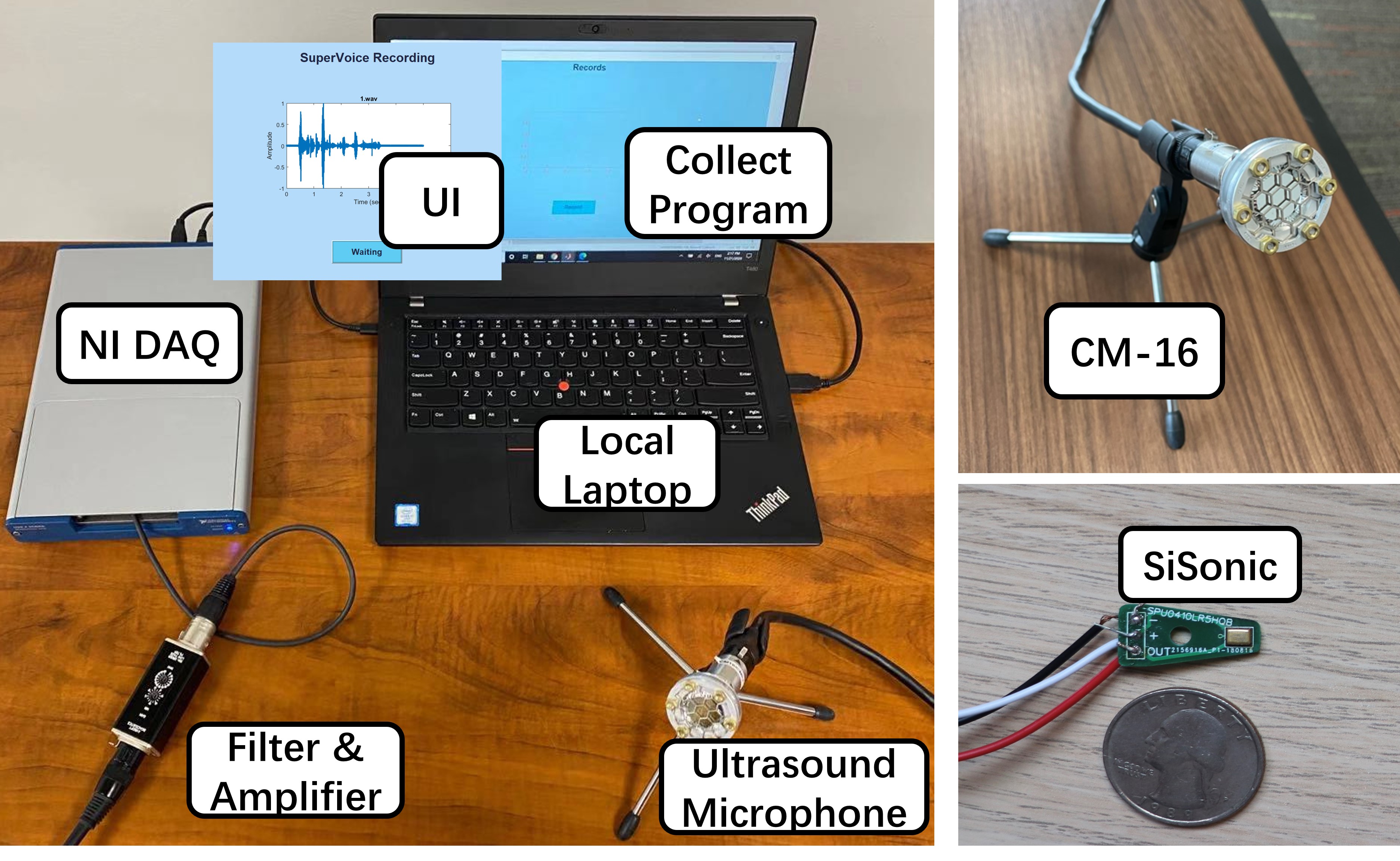}
\vspace{-10pt}
\caption{Dataset collection platform.}
\label{fig:platform}
\vspace{-15pt}
\end{figure}

\subsection{Speech Data Collection}
\label{sec:data_collect}
\noindent \textbf{Human Voice Recording.} The voice data in all the existing public datasets is collected using regular microphones with at most $48$ kHz sampling rate~\cite{garofolo1993timit, nagrani2017voxceleb} 
to record data within [0-24] kHz.

In order to investigate the high-frequency ultrasound components in the human speech, we collect our datasets for evaluation, including Voice-1, Voice-2, and Voice-3 datasets. The Voice-1 is collected by a high-end ultrasound microphone, Voice-2 is collected by regular microphones on various smartphones, and Voice-3 is collected by a low-end ultrasound microphone. \emph{In total, we collected 9,050 speech utterances from 127 volunteers, the data collection process and user study have been approved by our school's IRB board.}

\noindent\textbf{Dataset Collection Platform:} There are many options for the off-the-shelf ultrasound microphone (i.e., SiSonic SPU0410LR5H\_QB MESE~\cite{microphone} and Avisoft condenser microphone CM16~\cite{cm-16}). The first microphone can capture the ultrasound frequency band up to 96 kHz, which only requires a 1.5 V to 3.6 V power supply. The low power consumption and low cost (\$2/piece) make it suitable for most smartphones. The second microphone provides a more flat frequency response over the entire frequency band, allowing it to collect better-quality ultrasound recordings. 
For this reason, we deploy both SiSonic SPU0410LR5H\_QB microphone and Avisoft microphone for data collection. The microphone and data capturing equipment are displayed in Fig.~\ref{fig:platform}. For each participant, we informed them of the purpose of the experiment and then recorded their voice. The participants spoke facing forward to the microphone with a distance of 30 cm. Each participant was requested to speak 4 types of sentences, totaling 100 sentences. The instruction for the experiment and a part of the sentences can be found in Appendix~\ref{appen:dataset}.

\noindent\textbf{Voice-1:}
Voice-1 includes the voice data from 77 volunteers, totalling 7,700 utterances using 192 kHz sampling rate.
Among the 77 volunteers, most of them are college students with ages ranging from 18 to 56, including 38 males and 40 females. For detailed dataset information, please refer to website \textbf{\url{https://supervoiceapp.github.io}}. 

\noindent\textbf{Voice-2:}
Voice-2 is constructed by recording 25 sentences by 50 participants with different models of smartphones listed in Appendix~\ref{appen:dataset}. The smartphones' sampling rate is $48$ kHz.
As traditional speaker model leverages voice features below $8$ kHz, Voice-2 helps validate the effectiveness of high-frequency features within [$8$, $24$] kHz range recorded using different phones. In total, Voice-2 includes $1,250$ utterances with $48$ kHz sampling rate.

\noindent\textbf{Voice-3:}
Voice-3 includes 200 audios recorded from 20 participants. Different from Voice-1, 
we collect Voice-3 by the cheap SiSonic ultrasound microphone. Every participant read  a sentence twice in Common type (See Appendix~\ref{appen:dataset}), in total 10 audios per volunteer. The purpose of Voice-3 is to validate the performance of \supervoice with a cheap ultrasound microphone that can be integrated into smartphones~\cite{spu0410}.

\noindent \textbf{Spoofing Voice Dataset:}
We implement the spoofing attacks by replaying the voice data collected in Voice-1 using 5 playback devices and 2 recording devices, including 2 smartphones, 2 high-end commercial loudspeakers, and one ultrasonic speaker. To detect the replay attack, we deploy an ultrasound microphone to record the replayed spoofing audio. The purpose of this dataset is to comprehensively evaluate the capability of ultrasound components for liveness detection. 
\subsection{Performance Metrics}
The performance metrics we use for the SV task 
are \emph{False Acceptance Rate (FAR)}, \emph{False Reject Rate (FRR)}, and \emph{Equal Error Rate (EER)}. FAR represents the rate of \supervoice falsely accepting an unauthorized speaker, FRR is the rate of \supervoice rejecting a legitimate voice, and EER is the rate where the FRR and FAR are equal. We further use \emph{Classification Error Rate (CER)} to evaluate the speaker recognition (SR) performance, which is defined as the ratio of misclassified recordings versus the total recordings. For the user  study, we develop \supervoice as an end-to-end desktop application, and use \emph{Success Rate} to measure the percentage of successful attack defenses by \supervoice, i.e., the times of correct recognition of the voice owner
over the total number of attempts. 

\subsection{Speaker Verification Performance of Integrated Models}
\label{sec:sv_result}
To make a fair comparison with other existing speaker models, we reproduce all the models in Pytorch framework. We use the  Pytorch version 1.2.0 with Python version 3.6.9. The GE2E~\cite{wan2018generalized} and Void~\cite{ahmedvoid} are closed source, which we reproduce based on their descriptions. 
The GMM-UBM~\cite{reynolds2000speaker}, VGGVox~\cite{nagrani2020voxceleb} have open-source MATLAB codes, while SincNet~\cite{ravanelli2018speaker} and STFT+LCNN~\cite{lavrentyeva2017audio} are implemented with the Pytorch framework. 
All of the models are evaluated with the same datasets described in Section~\ref{sec:data_collect}. 

\noindent\textbf{Direct Ultrasound Integration:}
First, we conduct a performance evaluation of 4 popular SV models: GMM-UBM, SincNet, VGGVox, and GE2E. 
We follow each model's specification to configure the input and model parameters. Then, we evaluate their performance using the downsampled low frequency data ([0-8] kHz) and the original data ([0-96] kHz for \textbf{Voice-1} and [0-24] kHz for \textbf{Voice-2}). The performance comparison is presented in Table~\ref{tab:table1}. The performance with low-frequency data is relatively consistent with their reported results. When the high-frequency data is included in the modeling process, the performance of every model deteriorates significantly. This indicates that the high-frequency data cannot be directly used to distinguish among different speakers. 

\begin{table}[h]
    \centering
    \caption{EER performance (\%) comparison among GMM-UBM, SincNet, VGGVox, GE2E with two different datasets.}
    \vspace{-10pt}
    \label{tab:table1}
    \begin{tabular}{c|c|c|c|c}
    \hline
    Speaker Model& \multicolumn{2}{c|}{\textbf{Voice-1} } &
    \multicolumn{2}{c}{\textbf{Voice-2} } \\
    \cline{2-5}
     (kHz) & $[0-8]$ & $[0-96]$ & $[0-8]$ & $[0-24]$ \\
    \hline
      GMM-UBM  & 12.25 & 42.23 & 13.33 & 17.56 \\
      \hline
     SincNet  & 4.17 & 18.23 & 4.19 & 7.04  \\
     \hline
     VGGVox   & 4.64 & 16.63 & 4.66 & 6.75  \\
     \hline
     GE2E     & 4.98 & 19.15 & 4.97 & 6.96  \\
     \hline
    \end{tabular}
\end{table}

\begin{figure}[!t]
\centering
\subfigure[Performance improvement ]{\label{fig:boost}\includegraphics[width=40mm]{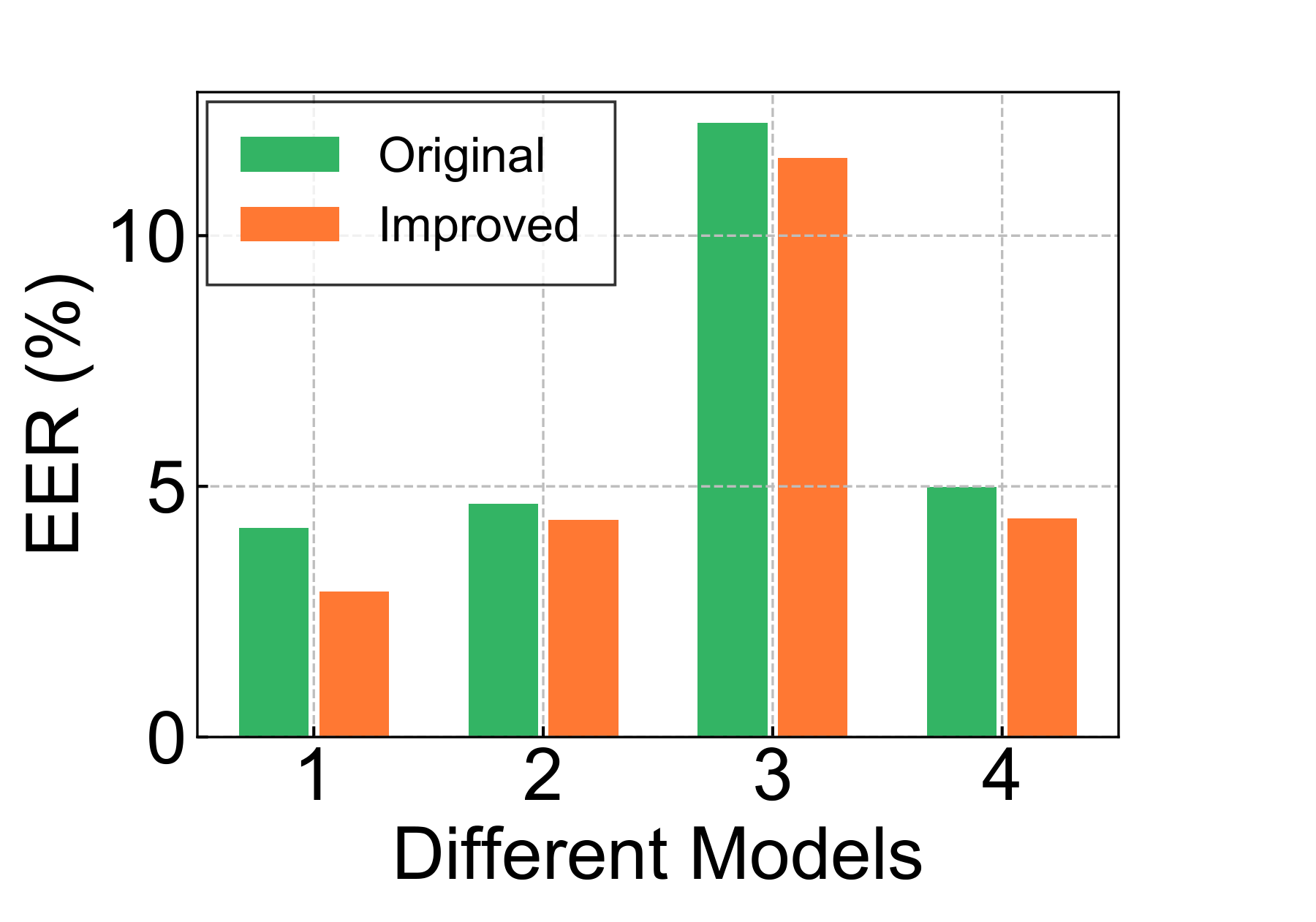}}
\subfigure[EER of \supervoice ]{\label{fig:eer}\includegraphics[width=40mm]{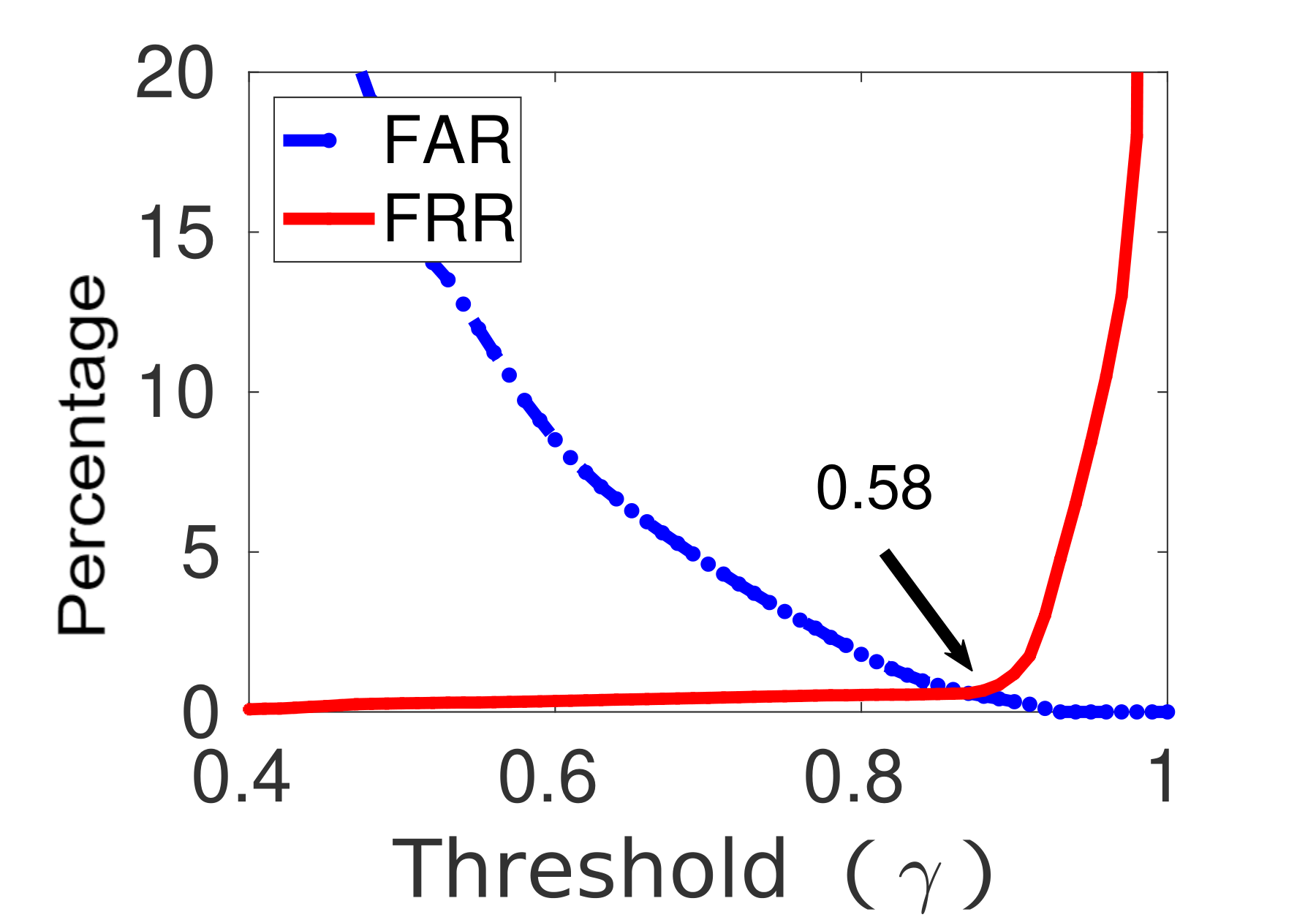}}
\vspace{-10pt}
\caption{Performance of: (a) ultrasound integration in existing models (1, 2, 3, and 4 represent SincNet, VGGVox, GMM-UBM, and GE2E models), tested on \textbf{Voice-2}; (b) ultrasound integration in \supervoice system, tested on \textbf{Voice-1}.}
\vspace{-5pt}
\label{fig:performance-ultrasound}
\end{figure}

\noindent\textbf{Improved Ultrasound Integration:}
For a better integration of the high-frequency data, we adopt the architecture of \supervoice: (1) using \emph{CNN-2} to handle high frequency data, and (2) replacing \emph{CNN-1} with the existing speaker models. To validate the efficiency of integrating high frequency data in the smartphones, we conduct an experiment with \textbf{Voice-2} ([0-24] kHz) and present the results in Fig.~\ref{fig:boost}. The green bar with rectangle pattern indicates the EER performance with the downsampled [0,8] kHz data, and the orange bar with cross pattern shows the performance with the addition of high-frequency features in the range of [8, 24] kHz that are extracted by \emph{CNN-2} and feature fusion technique. 
The results show that the EER of SincNet has a drop from 4.19\% to 2.89\%, and the EER of VGGVox decreased from 4.66\% to 4.12\%. 
Overall, the EER performance improvement is around 16.93\% on average with SincNet, VGGVox, and GE2E. For the GMM-UBM model, the EER performance has also improved slightly.
The results demonstrate \supervoice's transferability, i.e., it improves other SV models' performance by integrating the high-frequency feature embeddings. 
We then evaluate the performance of ultrasound integration in \supervoice using Voice-1 dataset. The result in Fig.~\ref{fig:eer} shows the FAR and FRR of \supervoice w.r.t. similarity threshold $\gamma$, and it indicates that the EER performance of \supervoice
is 0.58\%. 

\begin{table}[h]
    \caption{EER performance (\%) of \supervoice on different datasets}
    \label{tab:freqencyresult}
\centering
\vspace{-5pt}
\begin{tabular}{c|c|c|c|c|c|c}
\hline
\multirow{2}{*}{\supervoice} & \multicolumn{2}{c|}{\textbf{Voice-1}} & \multicolumn{2}{c|}{\textbf{Voice-2}}    & \multicolumn{2}{c}{\textbf{Voice-3}} \\
\cline{2-7}
                            & SV           & SR           & SV                    & SR                    & SV           & SR           \\
                            \hline
No HFE                      & 4.17         & 5.87         & 4.19                  & 6.74                  & 6.75         & 7.84         \\
\hline
{[}8-16{]} kHz              & 3.98         & 4.79         & 3.77                  & 4.87                  & 5.74         & 5.95         \\\hline
{[}8-24{]} kHz              & 2.89         & 2.27         & 3.21                  & 3.32                  & 3.45         & 4.51         \\\hline
\textbf{{[}8-48{]} kHz}              & \textbf{0.58}         & \textbf{1.61}         & - & - & \textbf{1.87}         & \textbf{3.01}         \\\hline
{[}8-96{]} kHz              & 5.79         & 7.31         & -                     & -                     & 9.52         & 14.2       \\ \hline
\end{tabular}

\end{table}

\begin{figure*}[!t]
\centering
\subfigure [t-SNE result]{\label{fig:tsne}\includegraphics[width=42mm]{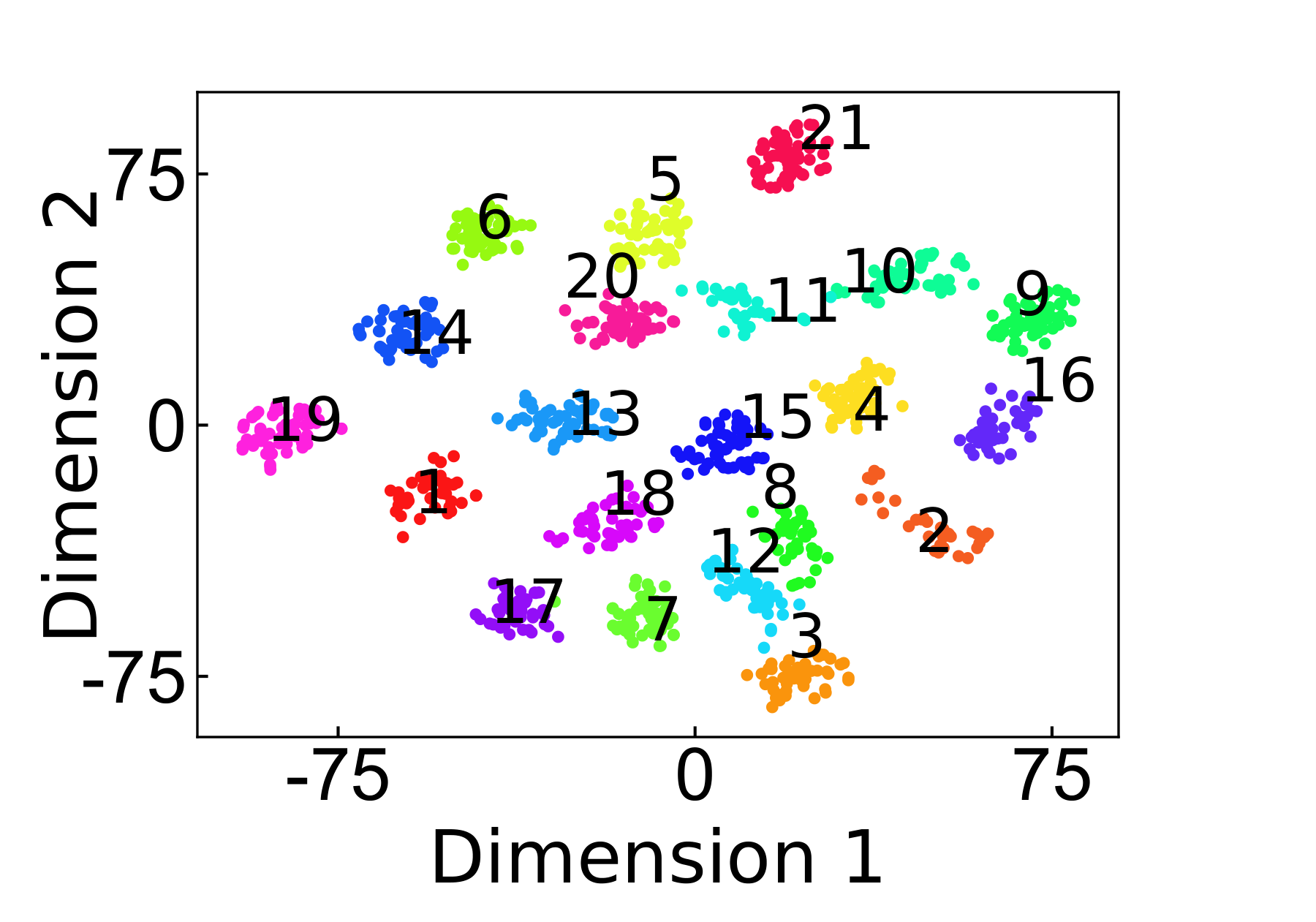}}
\subfigure[Distance vs. success rate ]{\label{fig:distance}\includegraphics[width=42mm]{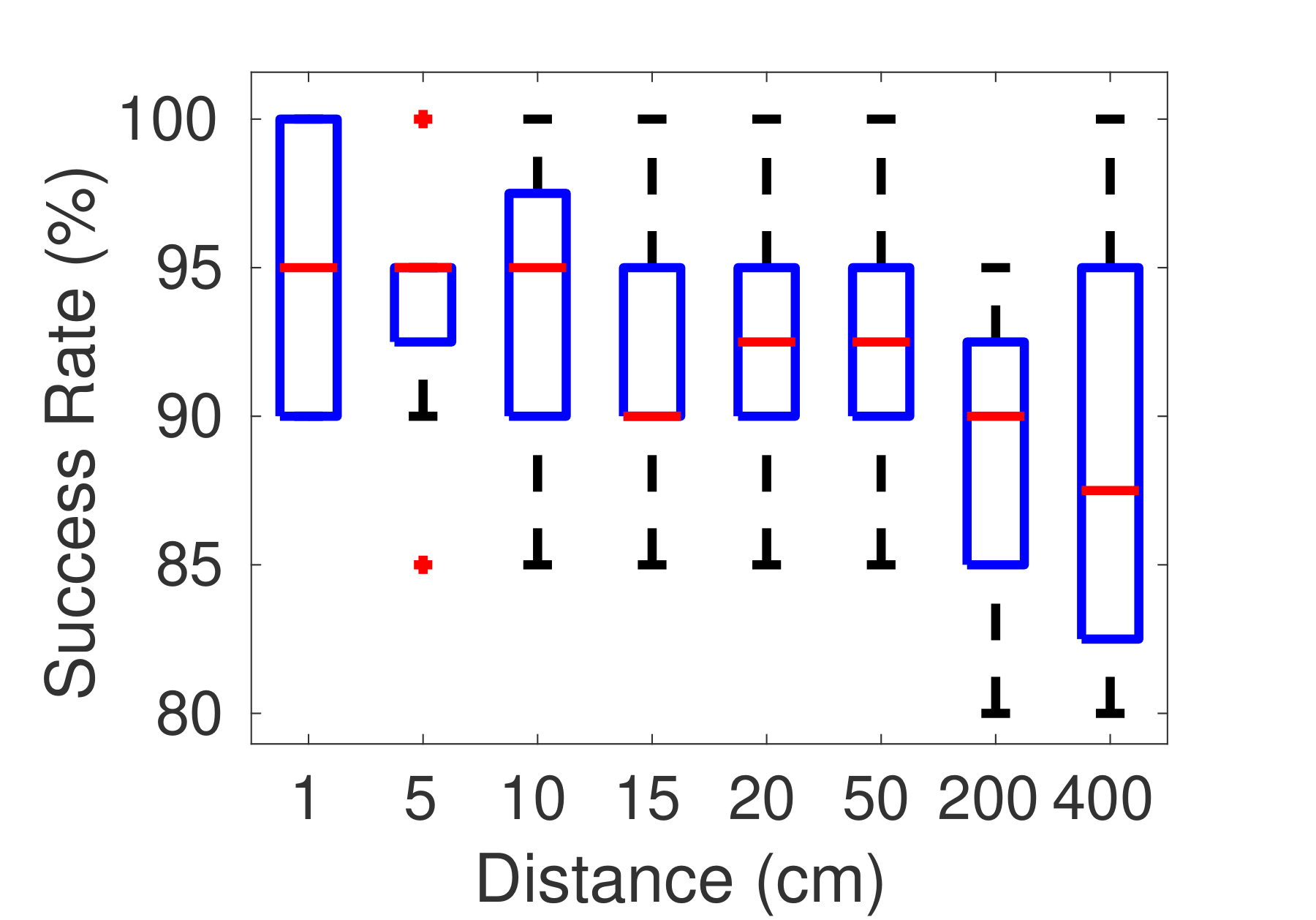}}
\subfigure[Angle vs. success rate ]{\label{fig:angle}\includegraphics[width=42mm]{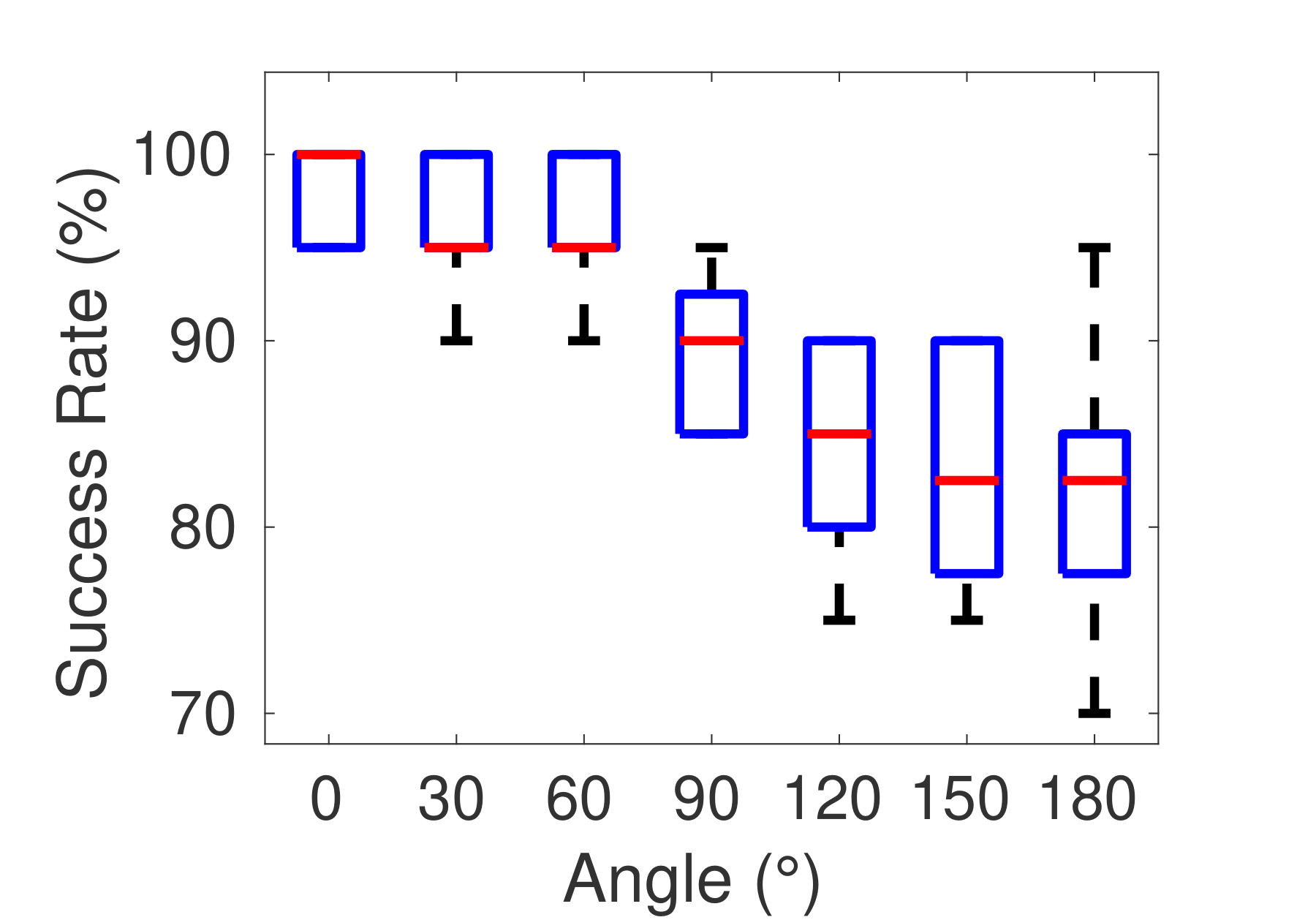}}
\subfigure[Durability vs. success rate ]{\label{fig:longe}\includegraphics[width=42mm]{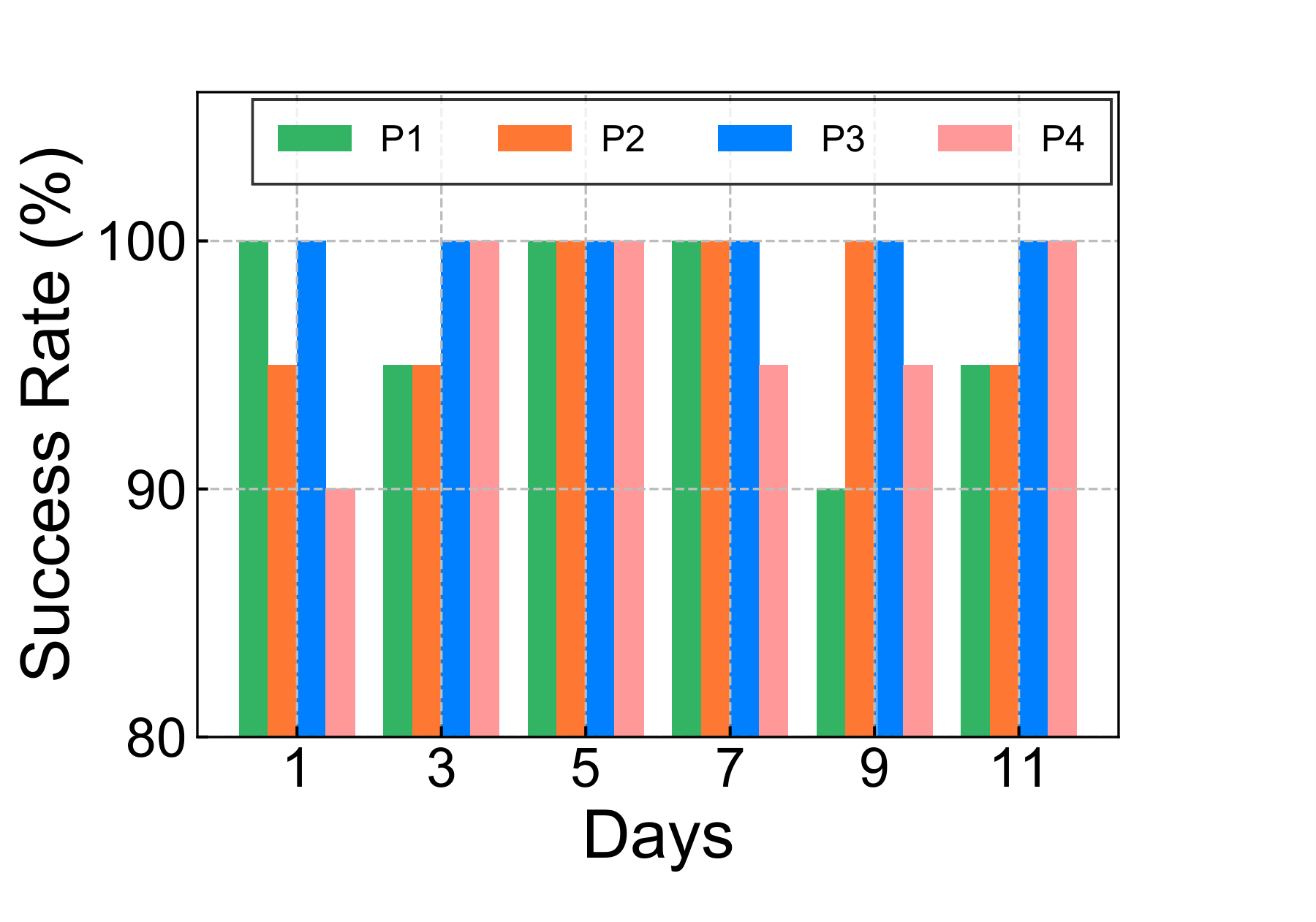}}
\vspace{-10pt}
\caption{The user study of \supervoice. }
\vspace{-15pt}
\label{fig:Experiment Settings}
\end{figure*}

\vspace{-7pt}
\subsection{Impact of Frequency Ranges}

Next, we evaluate the performance of \supervoice with different frequency ranges of the high-frequency data. Both SV (i.e., to verify the voice is from an authorized user) and speaker recognition (SR) (i.e., to recognize the voice of a specific authorized user) tasks are conducted to measure the EER and CER performance. 
The results in Table~\ref{tab:freqencyresult} show that \supervoice can achieve the EER performance of 0.58\% with Voice-1, the best among all the existing speaker models. It is noteworthy that the best models that tested on Voice-1 is SincNet, which has 4.17\% EER (see Table~\ref{tab:table1}). 
To further evaluate  \supervoice on smartphones with an affordable ultrasound microphone (e.g., SiSonic SPU0410LR5H), we evaluate the performance with the Voice-3 dataset. 
The results show that, even with low-end ultrasonic microphone, \supervoice achieves significant performance improvement. 

\emph{Remarkably, \supervoice improves the EER performance of the best SV model by 86.1\% (or 55.1\% with low-end microphone), via the incorporation of ultrasound frequency components.}
We also find that incorporating high-frequency features below $48$ kHz will produce better performance compared with the higher frequency range. Among all the configurations of frequency ranges, the range of $[8,48]$ kHz provides the best SV and SR performance in terms of EER and CER. 
Unsurprisingly, with a complete spectrum of $[0,96]$ kHz, both SV and SR performance degrades, as more indistinguishable noises are incorporated in the model to perplex the SR/SV tasks.

\subsection{\rev{User Study}}
Besides the benchmark evaluation \rev{presented in the previous} sections, we perform two \rev{user studies} to further test the effectiveness and robustness of our system.
(a) \textbf{Flexibility Study:} users use our system at home and make speeches from random positions; (b) \textbf{Longevity Study:} users use our system over a long time span. Fig.~\ref{fig:tsne} visualize the t-SNE result~\cite{maaten2008visualizing} of 20 participants in a 2D space, which clearly shows the 20 clusters of speakers. 

To conduct the user studies more efficiently, we develop an end-to-end \supervoice desktop application. 

\noindent\textbf{Flexibility Study:} We deploy the end-to-end application and ask 8 volunteers to enroll in the application. Once they are successfully enrolled, they are instructed to speak to the ultrasound microphone at different distances and angles to test the system's recognition performance. 
Each volunteer \rev{makes} 20 test attempts. \rev{Figs.}~\ref{fig:distance} and \ref{fig:angle} present the impact of distance and \rev{angle,} respectively. The results show that \supervoice reaches high success rate ($95-100\%$) within $50$ cm. \rev{Although} the success rate \rev{may drop} to $85\%$ beyond $50$ cm \rev{in the worst case}, \rev{the average accuracy at $400$ cm reaches $87.5\%$}. As for different angles, the recognition performance declines from $95\%$ to $85\%$ when the speaker is side facing the microphone. The \rev{performance degradation} is caused by \rev{a specific} characteristic of \rev{the} ultrasound microphone (\rev{i.e.,} CM-16 \rev{delivers} different gains at different angles according to its polar diagram).

\noindent\textbf{Longevity Study:} For the second user study, we test the longevity performance of our system by tracking the usage of 4 users over 11 days. The participants enroll their voices on the first day, and attempt 20 times per day to use \supervoice to identify their respective voices. As illustrated in Fig.~\ref{fig:longe}, the average success rate is more than $95\%$, which means less than 1 over 20 attempts failed.
In the end, we found no evidence of a degrading performance pattern over  time.

\subsection{Runtime Performance}
In this section, we compare the training time and testing time of \supervoice with SincNet, VGGVox, GMM-UBM, and GE2E models. The training time is the total time used to create a speaker model with the training pool of Voice-1, while the testing time represents the time spent to verify
an incoming utterance. 
\vspace{-5pt}

\begin{table}[h]
    \centering
    \caption{Runtime comparison.}    
    \label{tab:table2}
    \vspace{-5pt}
    \begin{tabular}{c c c}
    \hline
    \multirow{2}{*}{\textbf{\rev{Model}}} & \textbf{Training} & \textbf{Testing} \\
    & \textbf{time (sec.)} & \textbf{time (sec.)} \\
    \hline
       GMM-UBM  & 7,149 & 0.074 \\
     \hline
     VGGVox & 11,308 & 0.279 \\
     \hline
    GE2E  & 10,348 & 0.21 \\
     \hline
     SincNet  & 8,180 & 0.134 \\
        \hline
     \textbf{\supervoice}  & \textbf{8,413} & \textbf{0.120} \\

     \hline
    \end{tabular}
    \vspace{-10pt}
\end{table}

Table~\ref{tab:table2} presents the runtime result. Among all the models, the GMM-UBM model is the fastest in terms of training and testing time with the worst EER.
SincNet converges very fast during the training phase due to its special convolutional neural design, while the proposed \supervoice also delivers comparable training time. 
During the testing, \supervoice outperforms VGGVox and GE2E models due to its lightweight model with a small number of parameters. 
It is worth noting that introducing high-frequency features does not affect the testing speed. The results show that \supervoice could retain comparable runtime performance with enhanced speaker verification performance. 

\subsection{Liveness Detection Performance}
In this section, we conduct experiments to verify the performance of liveness detection described in Section~\ref{sec:live}. We prepare two types of recorders and 5 playback devices
to replay the recordings. For every speaker, we replay 20 audios at a fixed position (facing forward in 10cm) and volume ($60dB_{SPL}$). The defender uses the low-cost SiSonic ultrasonic microphone to monitor the replayed audios.

\begin{figure}[t]
\centering
\subfigure[Recorded by smartphone recorder ]{\label{fig:replay-samsungsource}\includegraphics[width=40mm]{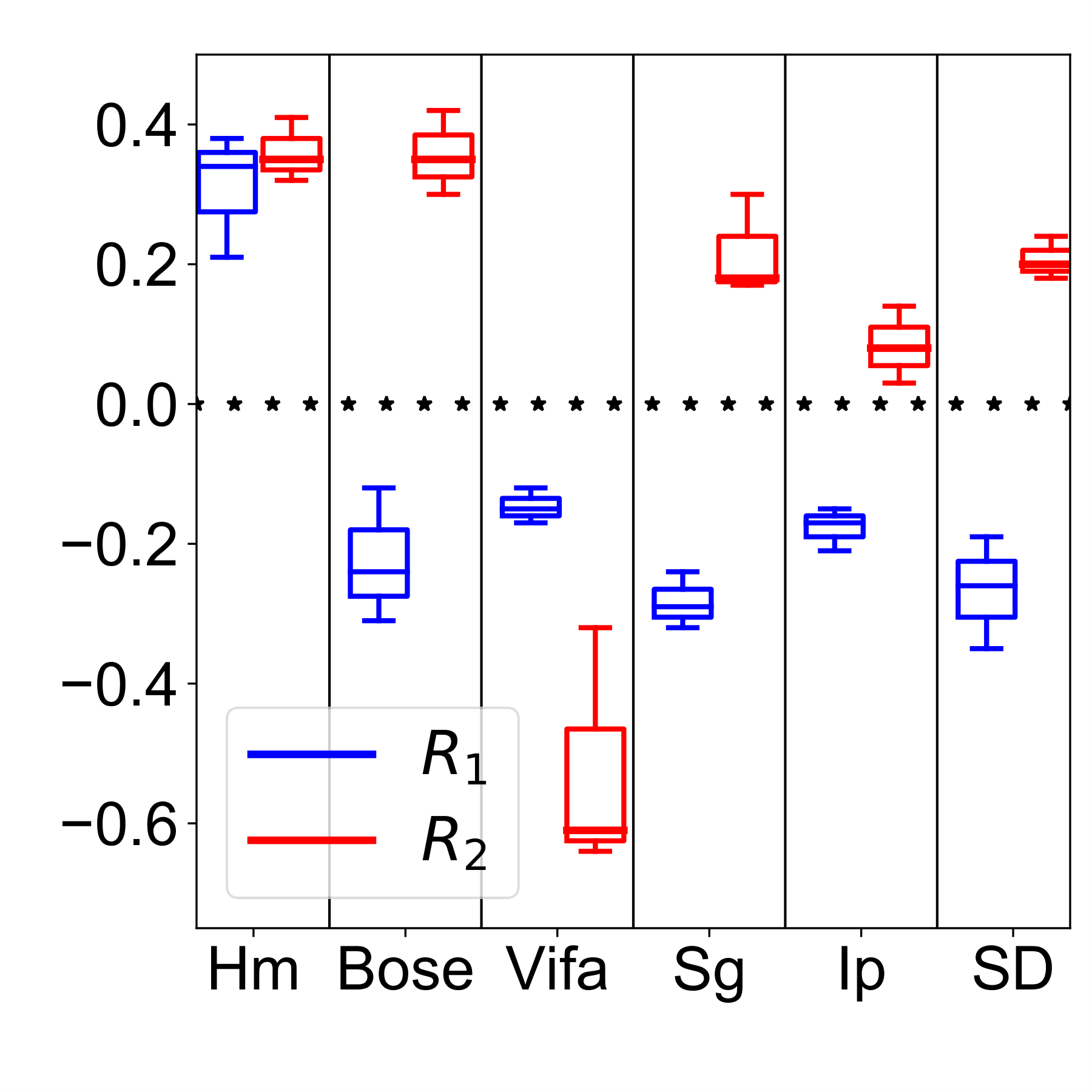}}
\subfigure[Recorded by ultrasound microphone ]{\label{fig:replay-ultrasource}\includegraphics[width=40mm]{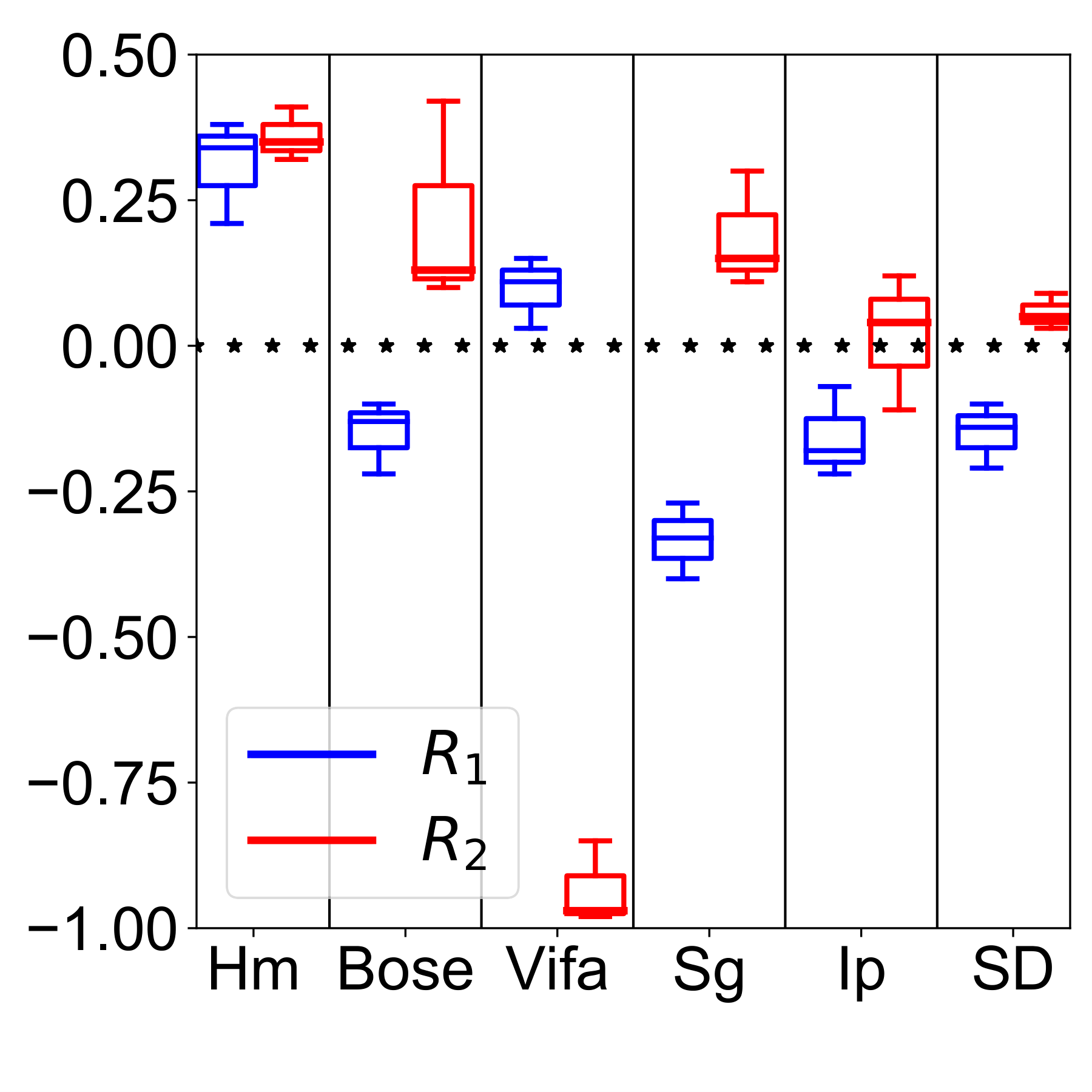}}
\vspace{-10pt}
\caption{Replay attacks}
\vspace{-10pt}
\label{fig:replay-attack}
\end{figure}

\noindent\textbf{Attackers Record with Common Recorder:}
We first replay audios that were recorded from a smartphone (Samsung S9). The boxplot in Fig.~\ref{fig:replay-samsungsource} demonstrates the results with different speakers. From left to right, we have Human genuine voice (Hm), Bose SoundTouch 10 speaker~\cite{bose}, Vifa ultrasonic speaker~\cite{vifa}, Samsung S9 phone (Sg), iPhone 12 (Ip), and SADA D6 speaker~\cite{sada}. The results show that all the replay devices present a negative $R_1$. This is attributed to the lack of HFE in the recorded audios by the smartphone. 
In contrast, the genuine human voices have positive $R_1$ and $R_2$, which is consistent with our analysis in Section~\ref{sec:live}. 
In the end, \supervoice achieves $0\%$ EER. 

\noindent\textbf{Attackers Record with Ultrasound Recorder:}
Now, we consider the attackers use a high-end ultrasonic microphone to record the victims' voices. 
We select 20 audio samples with 192 kHz sampling rate in Voice-1 as the source to replay them by 5 loudspeakers. The result in Fig.~\ref{fig:replay-ultrasource} shows that the commercial speakers still cannot produce any HFE, yielding all negative $R_1$. 
Moreover, a substantial gap exists between the genuine and replayed voice from any specific replay devices, which indicates that the liveness detection of \supervoice is robust against any attack devices. For the attacker with an ultrasonic speaker (Vifa), we observe a positive $R_1$. However, its negative $R_2$ signifies the low LFE. 
In the end, 
\supervoice again achieves 0\% EER, consistently confirmed by 200 attack attempts.

\begin{table}
    \caption{Liveness detection performance comparison.}
    \vspace{-5pt}
    \label{tab:table3}
    \centering
        \begin{tabular}{{p{3.1cm}p{1cm}p{1.7cm}p{1cm}}}
          \toprule[1.5pt]
          \textbf{Models} & \textbf{\# Feat.} & \textbf{Time (sec.)} & \textbf{EER(\%)} \\
          \midrule
          CQCC + GMM\cite{kinnunen2017asvspoof} & 14,020 & 0.159 & 12.08 \\
          LPC + CQCC + GMM & 14,026 & 0.173 & 13.74 \\
          STFT + LCNN\cite{lavrentyeva2017audio} & 84,770 &  0.321 & 8.8 \\
          Void\cite{ahmedvoid} & 97 & 0.103 & 11.6 \\
          \textbf{\supervoice} & \textbf{4} & \textbf{0.091} & \textbf{0} \\
      \bottomrule[1.5pt]
        \end{tabular}
        \vspace{-10pt}
\end{table}

\noindent \textbf{Defense Performance Comparison:}
Here, we compare the liveness detection performance with 4 state-of-the-art liveness or spoofing detection models. 
We first justify our reproductions by testing them on ASVSpoof~\cite{kinnunen2017asvspoof} dataset and all of them reach similar performance as they claimed. We then evaluate all the models using our spoofing dataset in terms of the number of features, average detection time, and the EER performance. 
Table~\ref{tab:table3} presents the liveness detection performance comparison results.
Among all the models, the STFT+LCNN model runs the slowest with the most number of features, while its EER performance is the best among the four models. 
Compared with the existing models, \supervoice only uses four accumulative power features in $R_1$ and $R_2$, and achieves the fastest runtime performance with 0\% EER. 
In consistent with the measured data in Fig.~\ref{fig:replay-attack}, which visualizes the manifest gap between genuine and spoofed sound, 
\supervoice achieves superior liveness detection performance in terms of both the runtime and EER performance for both the traditional loudspeakers and ultrasound speakers.

\section{Discussion}
\label{sec:discussion}

In this section, we discuss the limitations of \supervoice, the defense against the inaudible attacks\rev{, and} the future research directions.

\noindent \textbf{Commands Without Fricative Consonant\rev{:}} 
As mentioned before, we observed that some phonemes\rev{,} especially the fricative and stop consonants\rev{,} retain high energy above $20$ kHz. However, if a spoken sentence does not contain any fricatives, we may not \rev{be} able to find an energy spike in the spectrum. Fortunately, we observe the HFE from most of \emph{Non-fricative command} because the speaker always alters the air flow by their articulations, and this high-frequency component can be adopted by \supervoice as an extra feature for speaker verification. For those sentences that only include low-frequency energy (below 8 kHz), the low-frequency stream of our DNN architecture guarantees that \supervoice does not experience \rev{any} performance degradation with high-frequency features extracted from the non-fricative commands. 

\noindent\textbf{\rev{Long-Range} Speaker Verification\rev{:}}
In this paper, we assume \rev{that} the human speakers are \rev{within} a close distance \rev{from} the ultrasound microphone. Prior research found that
long range speaker verification is challenging mainly due to the reverberation of sound and attenuation of the acoustic energy~\cite{nandwana2018robust}. In \supervoice, 
the range of voice commands will affect the received power of both the low-frequency and high-frequency components, especially for the fricative and plosive consonants. A power amplifier may be able to address the power attenuation issue, and we plan to evaluate its effectiveness in \rev{a long-range} speaker verification in the future work.

\noindent\textbf{\supervoice on Smartphone\rev{:}} 
In our experiments, we run \textsc{SuperVoice} on a desktop with an ultrasound microphone. We experiment with smartphones supporting high sampling rate (i.e., $192$ kHz) to capture high-frequency 
voice components. Yet, we find that, due to the low-pass filter in the microphone system, all the frequency components above $24$ kHz \rev{have} been filtered out. 

One possible solution is to 
replace the microphone in the smartphone with the one supporting ultrasound frequency~\cite{microphone}, or use an external microphone that \rev{can be connected} to \rev{the} smartphone.
We also evaluate the performance of an external ultrasound microphone, i.e., Echo Meter Touch 2~\cite{ultramicro}, in capturing high-frequency components in voice signals. The external ultrasound microphone is attached to Samsung Galaxy S9 with a sampling rate of $256$ kHz. The results show that the voice data captured by external microphone can achieve similar SV performance as the standalone ultrasound microphone.

\noindent\textbf{Inaudible Attack Defense:}
The inaudible attacks leverage the non-linearity of \rev{a} microphone to perform \rev{an} inaudible command injection attack through ultrasonic speakers~\cite{zhang2017dolphinattack,yan2020surfingattack,roy2017backdoor}. The basic idea is to modulate the voice commands to ultrasound frequency band, and then transmit the modulated signal through an ultrasonic speakers. Due to the non-linearity of the regular microphone, the ultrasonic signal will shift frequency to \rev{the} audible frequency range in the microphone. As a result, the command can be perceived by voice-activated devices.

\begin{figure}[t]
\centering
\subfigure[{Inaudible attack on regular microphone} ]{\label{fig:attack1}\includegraphics[width=40mm]{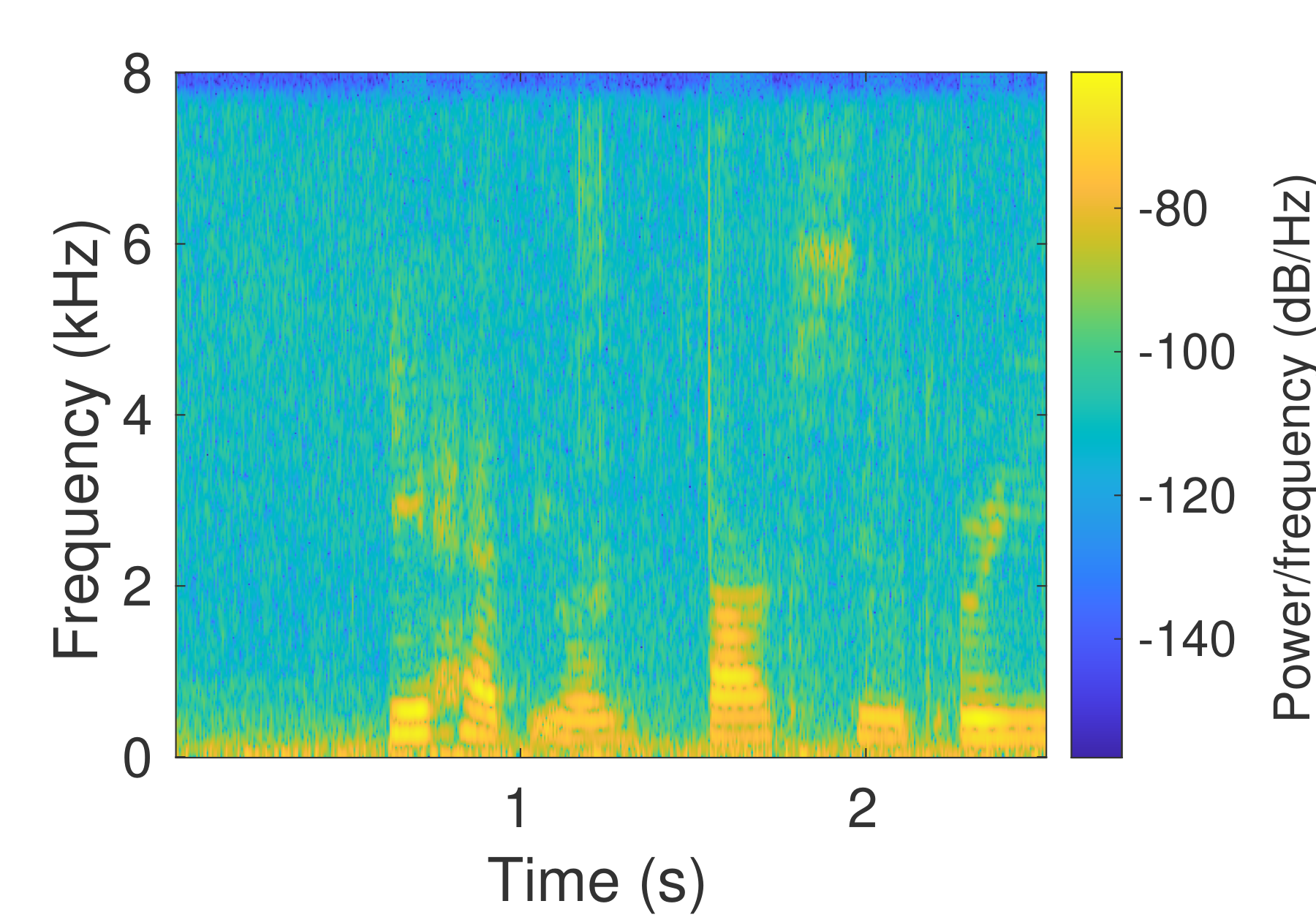}}
\subfigure[Inaudible attack captured by \supervoice ]{\label{fig:attack2}\includegraphics[width=40mm]{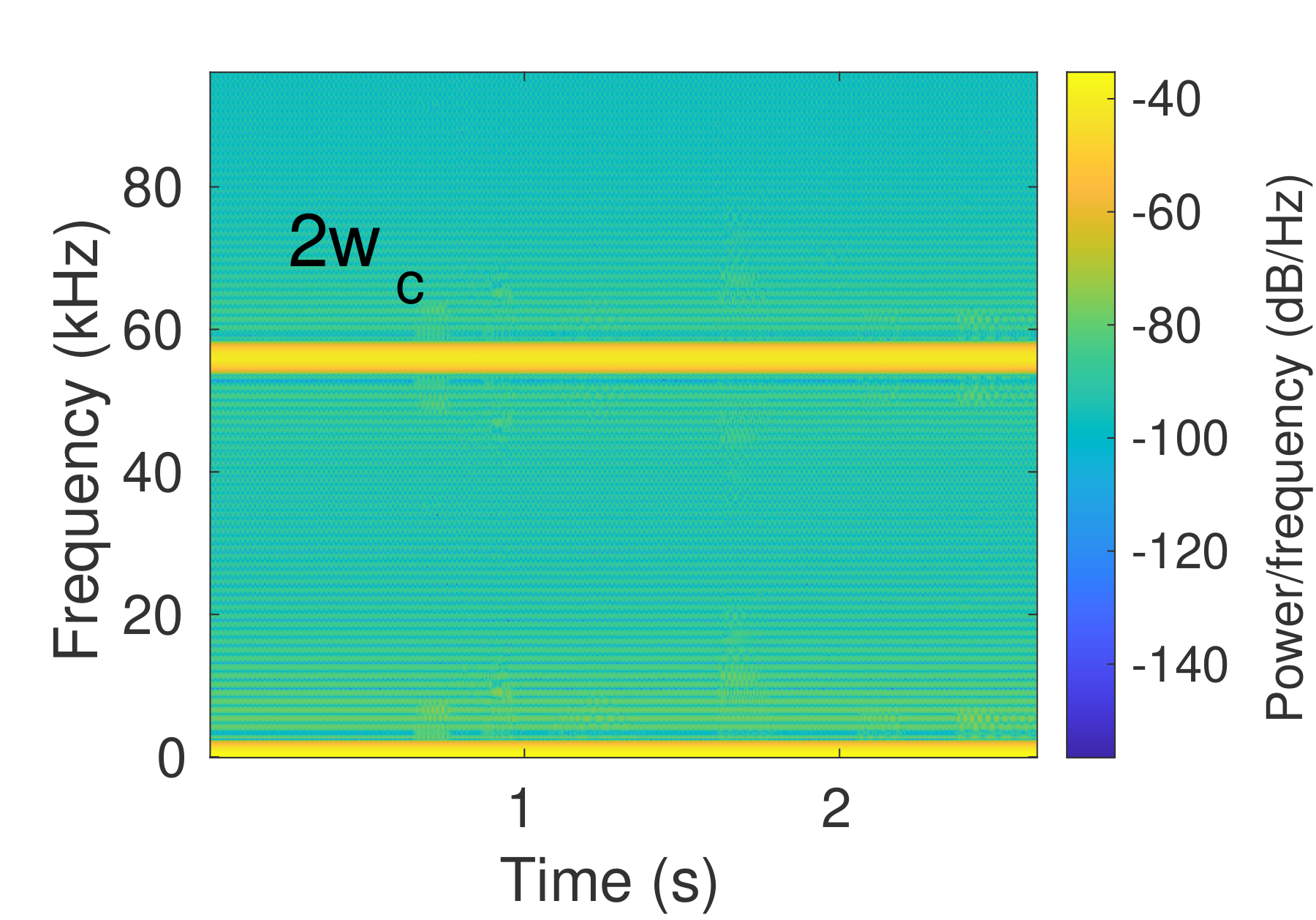}}
\vspace{-10pt}
\caption{The defense against inaudible attacks.}
\vspace{-10pt}
\end{figure}

We evaluate \supervoice's capability in detecting inaudible attacks. 
We use a voice command ``\emph{She had your dark suit in greasy wash water all year}" from Google TTS as \rev{a} legitimate signal. This command is modulated to the inaudible frequency at $w_c=28$ kHz. 
Figs.~\ref{fig:attack1} and \ref{fig:attack2} show the spectrogram of inaudible attack towards both \rev{a} regular microphone and \supervoice's ultrasound microphone. The 
regular microphone only captures frequency components \rev{in the range} $[0,8]$ kHz, while \supervoice can capture a $2w_c=56$ kHz component that can be used to immediately detect the inaudible attack. Therefore, \supervoice effectively defeats inaudible command injection attacks to voice assistants.

\section{Related Work}
\label{sec:related-work}

\noindent\textbf{Speaker Verification:} Prior studies have identified different voice features for speaker verification models. 
They use speech spectrum, speaker pitch and formants, and even raw audio waveforms as inputs~\cite{sainath2015learning, hoshen2015speech, kinnunen2010overview, ravanelli2018speaker, heigold2016end}, from which
 various voice features can be extracted, such as Filter Banks, MFCC (Mel-Frequency Cepstral Coefficients), LPC (Linear Prediction Coefficients), LPCC (Linear Prediction Cepstral Coefficients), or any combination of them~\cite{variani2014deep, snyder2017deep, chowdhury2019fusing}. 
With the voice features in hand, researchers further use GMM-UBM (Gaussian Mixture Model Universal Background Model)~\cite{campbell2006support},  JPA (Joint Factor Analysis)~\cite{kenny2005joint}, and neural networks~\cite{dehak2010front, heigold2016end,variani2014deep,chen2015locally, nagrani2017voxceleb, chung2018voxceleb2, nagrani2020voxceleb, wan2018generalized, ravanelli2018speaker} to generate speaker models. Based on the speaker models, several classifiers such as support vector machine (SVM)~\cite{campbell2006support, wan2005speaker}, cosine similarity~\cite{heigold2016end, shum2010unsupervised}, and PLDA (Probabilistic Linear Discriminant Analysis)~\cite{kenny2013plda, dehak2010front} have been employed to make (mostly probabilistic) SV decisions. 

\noindent\textbf{Spoofing Detection:} Existing spoofing detection solutions explore both non-vocal and vocal physical parameters of the human speaker to differentiate between human voice and spoofed sound. Among the approaches that use non-vocal physical parameters is VoiceLive~\cite{zhang2016voicelive}, which leverages smartphone's two microphones to capture the difference in the time of arrival (ToA) of phonemes to identify spoofing attacks.
Although VoiceLive does not require heavy computation, the detection accuracy largely depends on the distance between the speaker and the microphones. 
VoiceGesture~\cite{zhang2017hearing} performs liveness detection by identifying human gestures from the microphone-sensed Doppler shifts. 
VoiceGesture is designed for smartphones, which cannot be directly applied for voice-controlled IoT devices due to its stringent requirement on the positions of devices' microphones. 
Recently, WiVo~\cite{meng2018wivo} uses wireless sensing to detect lip movements associated with the syllables in the voice command, which requires to place wireless antennas very close to the speaker. Tom et al.~\cite{tom2018end} achieve a significant reduction of errors in replay attack detection
using an adaptation of the ResNet-18 model. Void~\cite{ahmedvoid} proposes a set of lightweight features in the audible spectrum to distinguish the voice source, and achieves low latency while maintaining relatively high detection accuracy. CaField~\cite{yan2019catcher} leverages the sound field characteristics to detect loudspeaker-based spoofing attacks.

Although the existing studies have achieved remarkable success in utilizing audible information of human voice, they either suffer from low accuracy on text-independent verification task, or require substantial computational resource usage. Different from all the previous approaches, the proposed \supervoice aims to provide a more accurate and realistic SV solution using the high-frequency ultrasound components in the human voice.

\noindent\textbf{Speaker Recognition Using High Frequency:} The utilization of high-frequency components of human voice for speaker recognition has been studied before~\cite{monson2014perceptual,hayakawa1994text,hayakawa1995influence}. These studies, however, are lacking of crucial technical details necessary for designing a contemporary high-performance text-independent SV system.
\section{Conclusion}
\label{sec:conclusion}
In this paper, we initiate an exploration on 
the underexplored ultrasound voice components in human speech, and we find that they can be used to enhance the performance of speaker verification and liveness detection. We design a speaker verification system, \supervoice, to show the strength of ultrasound frequency components in the speaker models. Specifically, we design a two-stream DNN structure to fuse the low-frequency and high-frequency features. \supervoice significantly improves the speaker verification and liveness detection performance in comparison with the existing models. We further demonstrate the possibility of integrating ultrasound frequency features in the existing models to enhance their verification performance. \supervoice is accurate, lightweight, and secure, which can be integrated into smartphones with a modification of smartphone's microphone component. 

\section*{ACKNOWLEDGEMENT}
\label{sec:ack}

We would like to thank the anonymous reviewers
for providing valuable feedback on our work. This work was supported in part by National Science Foundation grants CNS-1950171
and CCF-2007159.
Parts of the research reported in this publication was supported by the National Institute on Deafness and Other Communication Disorders, grant number R01DC012315. The content is solely the responsibility of the author and does not necessarily represent the official views of the National Science Foundation and National Institutes of Health.

\bibliographystyle{ACM-Reference-Format}
\bibliography{simple-base}

\appendix
\appendix
\section*{Appendix}

\section{Signal Processing}
\label{sec:sp}

The Signal Processing module takes raw voice signals as input. 
It first preprocesses the raw signals to remove the recorded silence. Then, it extracts ultrasound features from the original audio data, and downsamples the very high-quality audio to separate and extract low-frequency features.

\vspace{5pt}
\noindent\textbf{Signal Preprocessing:} The signal preprocessing removes the non-speech portions of the audio. Specifically, the silence periods within the speech are cropped to produce a clean speech signal. The original voice signal $s(t)$ is first segmented into $K$ frames $s_{i}(t)$, $i\in [1, K]$ (typically a segmentation length $K$ is 800 samples for $16$ kHz data). To crop the silence periods in $s_{i}(t)$, we initialize a \emph{silence period counter} $f(s_i(t))$ to count the number of contiguous silence periods: 
  \begin{equation}
    f(s_i(t)) =
    \begin{cases}
      0,  & \text{if  }~~~~\overline{(s_i(t))^2} > \theta \\
      f(s_i(t)) + 1,  & \text{if  }~~~~\overline{(s_i(t))^2} \le \theta,
          \label{eq:silence}
    \end{cases}
\end{equation}
$$\theta = min(\overline{(s_i(t))^2}) + (max(\overline{(s_i(t))^2}) -  min(\overline{(s_i(t))^2})) * \theta_{ratio}.$$
If the mean square power $\overline{(s_i(t))^2}$ of the current segment $s_{i}(t)$ is less than a pre-defined threshold $\theta$, this segment is marked as a silence period, and $f(s_i(t))$ will be incremented. We configure a constant number $c$ to indicate the toleration of contiguous silence frames. Segment $s_{i}(t)$ will be discarded if $f(s_i(t)) > c$, i.e., if all the previous $c$ frames are regarded as silence periods. The threshold in Eq.~(\ref{eq:silence}) depends on the power of the current frame and a constant value $\theta_{ratio}$, which specifies the percentile of the power distribution that we consider as background noise. Here, we empirically set the $\theta_{ratio}$ as $0.25$. The tolerance number $c$ is set to $25$, which means that if there are contiguous $25$ frames below the threshold, the upcoming frames will be discarded until a higher power frame arrives, where the higher power frame refers to a frame with its average power exceeding $\theta$. 

\vspace{5pt}
\noindent\textbf{Ultrasound Feature Extraction:} After the removal of the silence periods, we obtain the clean speech $s'(t)$, and then we perform STFT to convert the sound waveform into STFT spectrogram, which is a common approach adopted by the existing SV systems~\cite{nagrani2020voxceleb, heigold2016end, wan2005speaker}. 
Specifically, $s'(t)$ is first segmented into time frames of 2,048 samples, and then STFT results are computed using a Hann window with $\sfrac{1}{4}$ window length. 
We convert the amplitude spectrum to dB-scale. 
The FFT length is set as 2,048, while the sampling rate is $192$ kHz, and the frequency resolution is $192$ kHz$/2,048 = 93.75$ Hz. 
In the end, the feature extraction produces a matrix of STFT features with the size of $1,024 \times L$, and $L$ can be expressed as follows:
\begin{equation}
L = \frac{len(s'(t)) - win_{STFT}}{hop_{STFT}} + 1, 
\label{eq:window}
\end{equation}
where  $win_{STFT} = 2,048$, $hop_{STFT} = 512$. The STFT spectrogram can be divided into different frequency ranges. The ultrasound components between 24-60 kHz are used for liveness detection, while the components across the range of 8-96 kHz are adopted for speaker verification.

\noindent\textbf{Downsampling:} 
Besides the high-frequency components, we also extract the low frequency components as a separate input to the SV system. 
To do that, we perform a downsampling procedure using FFmpeg~\cite{ffmpeg} tool to reduce the sampling rate of clean speech $s'(t)$ from $192$ to $16$ kHz, and therefore obtain low frequency audio ranging in the interval $[0,8]$ kHz. 
The downsampled data contains the essential low-frequency components for SV~\cite{ganchev2005comparative, villalba2011preventing}. 
\section{Neural Network Implementation Details}
\label{appendix:neural}

\noindent\textbf{CNN-1:} The CNNs are utilized to extract low-frequency features. As the input of low-frequency feature is raw signals, we choose the Sinc-based convolutional layer~\cite{ravanelli2018speaker}. The Sinc-based convolutional network has demonstrated high efficiency and accuracy on speaker verification and recognition \rev{tasks} because it leverages interpretable filters with less network parameters. In our implementation, Sinc-based convolutional layer uses 80 filters with length \rev{$L=251$}.
The Sync layer is followed by 2 standard one-dimensional convolutional layers, both using 60 filters of length 5. We also apply a layer \rev{of} normalization for all the convolutional layers, and use the LeakyRelu function at every hidden layer to ensure the non-linearity of the CNNs model. The output dimension of the CNNs is 6420$\times$1.

\noindent\textbf{CNN-2:} This network takes high frequency cropped matrix as \rev{input} and \rev{generates} a high level feature of the high-frequency components. It contains 8 layers, the first two layers include 64 F-Filters with size of 9$\times$1 and dilated by 1$\times$1 and 2$\times$1 respectively. Followed by two horizontal layers, it contains 64 T-Filters with size of   1$\times$9 and dilated by 1$\times$1 and 1$\times$2. The next 4 layers are composed by square F/T filters, each layer has 48 filters with size of 5$\times$5, and dilated in various scale 2$\times$2, 4$\times$4, 8$\times$8 and 16$\times$16. The 8 layers are followed by an average pooling layer that produces a feature map with 512$\times$1 dimension. 

\noindent\textbf{NN:} The last NN model is used to produce speaker embeddings using the fused features from CNNs-1 and CNNs-2.
It is a fully connected layer that takes features with a dimension of 6420+512$\times$1 and output 2048$\times$1 dimension as the speaker embedding. Note that during the training procedure, the NN model also contains another fully connected layer and a softmax classifier, such that it is 
able to map 2048$\times$1 dimension to $class\times$1 to process the speaker recognition task, where $class$ is the number of speakers in the training pool. 

\noindent\textbf{Hyperparameters:} In the training stage, we use RMSprop optimizer, with a learning rate $lr = 0.001$, $\alpha=0.95$, $\epsilon=10^{-7}$, and the default minibatch is 128.

\section{Feature Alignment}\label{appen:feature}
One of the key challenges in \supervoice is to generate the input features from the low-frequency raw audio and high-frequency spectrum simultaneously. For this purpose, the CNN-1 and CNN-2 models require the aligned inputs of different formats (i.e., raw audio versus spectrum data). 
Given an incoming audio, we use one window to generate low-frequency raw audio frames, and another window to collect the windowed STFT spectrum data to be fit into the models. 
The window used for CNN-2 model matches the low-frequency window of the CNN model. In this paper, we set the window size of the low-frequency data $L_{win}$ to $200$ ms, corresponding to $200$ ms $\times$ $16$ kHz = $3,200$ samples, and the hop size $L_{hop}$ to 10 ms with $10$ ms $\times$ $16$ kHz = $160$ samples.
Then, the high-frequency window size $H_{win}$ and hop size $H_{hop}$ can be computed as follows: 
\begin{align}
\begin{aligned}
H_{win} &= L_{win} \times \alpha / hop_{STFT},\\
H_{hop} &= L_{hop} \times \alpha / hop_{STFT}, 
\end{aligned} \label{eq:slidewindow}
\end{align}
where $hop_{STFT}$ is the hop length of STFT spectrum (i.e., $512$ in this paper), and $\alpha$ is the ratio of high and low sampling rates. Here, the high and low sampling rates are $192$ kHz and $16$ kHz, respectively, i.e., $\alpha=12$. 
$hop_{STFT}$ is used to convert the number of samples in the low-frequency components into the number of columns in the STFT matrix of the corresponding high-frequency components. 
This window alignment scheme guarantees that these two types of features corresponding to the same original audio input are generated at the same time. Therefore, different types of input features, including both the low-frequency voice feature and ultrasound voice feature, can be fused together to form a coherent feature vector. 

\section{Dataset Collection and Composition}\label{appen:dataset}

\noindent\textbf{Dataset Collection Routine:}
To formulate our ultrasound dataset (e.g. Voice-1 and Voice-2), we recruit volunteers as many as we can by spreading information via social media, public Facebook page, and  posters on information boards. We provide each volunteer \$5 gift card for compensation. Here are the procedures we follow for each volunteer to conduct our experiment.

\textbf{Step 1:} Schedule a remote instruction session with each volunteer.

\textbf{Step 2:} Prepare scripts for the volunteer and disinfection on microphones and door handles.

\textbf{Step 3:} Instruct the volunteer outside the data collection room (e.g. getting familiar with the user interface). 

\textbf{Step 4:} Register the basic information of the volunteer.

\textbf{Step 5:} Conduct the voice recording experiment. 

\vspace{5pt}
\noindent\textbf{Types of Sentences:}
There are 4 types of sentences we designed.

\noindent\textbf{1. Common type:}
The common type contains 5 sentences, two of which are from the dialect set of TIMIT dataset~\cite{timit}, 
\vfill\eject
\noindent while the other three sentences are ``This is a test", ``Please identify me", and ``OK, please recognize my voice". 
For this type of sentence, each participant will speak every sentence twice, resulting in 10 recordings. 

\noindent\textbf{2. Compact type:}
The compact type has 200 sentences in total. They are randomly chosen from the TIMIT dataset, and designed to provide a good coverage of phonemes, with extra occurrences of phonetic contexts that are considered either difficult or of particular interest. Each participant reads 40 sentences from this type.

\noindent\textbf{3. Fricative command:}
The fricative commands are collected from the website \url{ok-google.io}, which provides commonly used commands on Google Assistant. We select 50 sentences that have fricative phonemes, e.g., ``find my phone", ``turn on my lights". Each speaker randomly picks 25 sentences from this type of commands.

\noindent\textbf{4. Non-fricative command:}
Similar to the fricative commands, the non-fricative commands are also collected from \url{ok-google.io}, which do not contain any fricatives. We randomly pick 25 sentences and ask every participant to read 25 sentences from this type.

\begin{table}[h]
    \centering
    \vspace{-10pt}
    \caption{Training and testing pools (\# of sentences).}
    \vspace{-10pt}
    \label{tab:speakers}
    \begin{tabular}{c|c|c|c|c}
    \hline
    \multirow{2}{*}{\textbf{Dataset}} & \multicolumn{2}{c|}{\textbf{Training pool}} &
    \multicolumn{2}{c}{\textbf{Testing pool}} \\
    \cline{2-5}
    & \textit{Training} & \textit{Valid.} & \textit{Enroll} & \textit{Verif.} \\
    \hline
      Voice-1  & 2,400 & 600 & 144 & 4656 \\ \hline
      Voice-2  & - & - & 75 & 1,175 \\\hline
      Voice-3  & - & - & 60 & 140 \\\hline
      \multicolumn{5}{l}{
    \vspace{-15pt}} \\
    \end{tabular}
\end{table}
Table~\ref{tab:speakers} shows the breakdown of training and testing pools for Voice-1 and Voice-2. For Voice-1, 30 participants are randomly selected into the training pool, and their recordings are then divided into the training and validation sets as described in Table~\ref{tab:speakers}. For participants in the training pool, their audios in the training set are used to train the \supervoice model, while audios in validation set are applied to validate the training process every 10 epochs. As for those in testing pool, the recordings are split into the enrollment and verification sets. In the testing procedure, \supervoice loads the trained model and uses the enrollment utterances to generate the speakers' embedding, and then it produces the embedding and calculates the similarity to make SV decision (i.e., accept or reject). All participants' voice data in Voice-2 are only used for testing. 
For Voice-2, we used 7 smartphones: iPhone X, iPhone XS, iPhone 8, iPhone Pro Max (Apple), Mi 9 (Xiaomi), and Mate 20 (Huawei). For Voice-3, we recruit 20 volunteers to speak 10 sentences. Then, we use 3 audios from every speaker as enrollment, and the rest of audios are used to verify the enrolled volunteers.

\end{document}